\documentclass[12pt,preprint]{aastex62} 
\def\lea{\mathrel{<\kern-1.0em\lower0.9ex\hbox{$\sim$}}} \def\gea{\mathrel{>\kern-1.0em\lower0.9ex\hbox{$\sim$}}} \newcommand{\lta}{{\>\rlap{\raise2pt\hbox{$<$}}\lower3pt\hbox{$\sim$}\>}} \newcommand{\gta}{{\>\rlap{\raise2pt\hbox{$>$}}\lower3pt\hbox{$\sim$}\>}}

\begin{document}

\title{The Fraction of Stars That Form in Clusters in Different Galaxies}

\correspondingauthor{Rupali Chandar}
\email{Rupali.Chandar@utoledo.edu}
\author{Rupali Chandar}
\affil{Department of Physics \& Astronomy, University of Toledo, Toledo, OH 43606}
\author{S. Michael Fall}
\affil{Space Telescope Science Institute, Baltimore, MD, 21218 USA}
\author{Bradley C. Whitmore}
\affil{Space Telescope Science Institute, Baltimore, MD, 21218 USA}
\author{Alexander J. Mulia}
\affil{Department of Physics \& Astronomy, University of Toledo, Toledo, OH 43606}

\begin{abstract}

We estimate the fraction of stars that form in compact clusters (bound and unbound), $\Gamma_F$, 
in  a diverse sample of eight star-forming galaxies, including two irregulars, two dwarf starbursts, two spirals, and two mergers.
The average value for our sample is $\Gamma_F\approx24\pm9$\%.
We also calculate the fraction of stars in clusters that survive to ages between $\tau_1$ and $\tau_2$, denoted by $\Gamma_S(\tau_1,\tau_2)$, and find $\Gamma_S(10,100)=4.6\pm2.5$\% and $\Gamma_S(100,400)=2.4\pm1.1$\%, significantly lower than $\Gamma_F$ for the same galaxies.
We do not find any systematic trends in $\Gamma_F$ or $\Gamma_S$ with the star formation rate (SFR), the SFR per unit area ($\Sigma_{SFR}$), or the surface density of molecular gas ($\Sigma_{H_2}$) within the host galaxy.
Our results are consistent with those found previously from the CMF$/$SFR statistic (where CMF is the cluster mass function), and with the quasi-universal model in which clusters in different galaxies form and disrupt in similar ways.
Our results, however, contradict many previous claims that the fraction of stars in bound clusters increases strongly with $\Sigma_{SFR}$ and $\Sigma_{H_2}$.
We find that the previously reported trends are largely driven by comparisons that mixed $\Gamma_F \approx \Gamma_S(0,10)$ and $\Gamma_S(10,100)$, where $\Gamma_S(0,10)$
was systematically used for galaxies with higher $\Sigma_{SFR}$ and $\Sigma_{H_2}$, and $\Gamma_S(10,100)$ for galaxies with lower $\Sigma_{SFR}$ and $\Sigma_{H_2}$.

\end{abstract}

\keywords{galaxies: star clusters:general --- stars: formation}

\section{Introduction}

Stars and star clusters have a common origin in the dense regions of molecular clouds (e.g., Lada \& Lada 2003; McKee \& Ostriker 2007).
It has long been known that the stellar initial mass function (IMF)---a direct product of star formation processes---is similar among different galaxies, possibly even ``universal'' (Bastian et al. 2010).
The common origin of stars and clusters suggests that there may also be similarities among cluster populations in different galaxies.
%The similarity of these distributions suggests that the formation and distribution of clusters are governed by ``quasi-universal'' processes.
We have found evidence for such similarities in the mass and age distributions of clusters, $\psi(M)\equiv dN/dM$ and $\chi(\tau)\equiv dN/d\tau$.
Both distributions can be represented by power laws, $\psi(M) \propto M^{\beta}$ and $\chi(\tau) \propto \tau^{\gamma}$, with exponents that are approximately (but not exactly) the same in different galaxies: $\beta \approx -2$ and $\gamma\approx-0.7$ (Fall \& Chandar 2012; see also Whitmore et al. 2007).
The similarity of these distributions suggests that the formation and disruption of clusters are governed by ``quasi-universal'' processes.

We showed recently that the mass functions of young clusters (with ages $\tau < 10^7$~yr), when divided by the star formation rate (SFR), are also similar among different galaxies (Chandar et al. 2015, hereafter CFW15; Mulia et al. 2016).
In our sample of 8 galaxies, the amplitude of the cluster mass function (CMF) and the SFR vary by factors $\sim10^3$, while their ratio (CMF$/$SFR) varies by less than a factor of two.
Moreover, we find no significant correlations between the CMF$/$SFR statistic and the other properties of the galaxies.
These results mean that the rates of star and cluster formation are essentially proportional to each other---another sign of quasi-universality.

The CMF$/$SFR statistic is closely related to another important quantity $\Gamma$, the fraction of stars that form in compact clusters.
Indeed, CMF$/$SFR and $\Gamma$ are proportional to each other (as we show in Section~2).
$\Gamma$ has figured prominently in several recent studies of cluster populations (e.g., Bastian 2008; Goddard et al. 2010; Kruijssen 2012, Adamo et al. 2015; Johnson et al. 2016).
It is usually defined as the fraction of stars that form in gravitationally bound clusters, i.e., those with negative total energy (kinetic plus potential).
However, since in practice the binding energies of clusters are never measured or even estimated, $\Gamma$, despite its putative definition, must be regarded as the fraction of stars that form in all compact clusters, both bound and unbound.
The most striking claim about $\Gamma$ from recent studies is that it increases systematically with $\Sigma_{SFR}$ and $\Sigma_{H_2}$, the mean surface densities of SFR and molecular gas in galaxies.
Thus, there is a stark discrepancy between our findings for CMF$/$SFR,
which shows no dependence on properties of the host galaxies, and the findings of others for $\Gamma$.
A major goal of this paper is to resolve this discrepancy.

One of our key ideas is illustrated in Figure~1, showing the increase of $\Gamma$ with $\Sigma_{SFR}$ based on data from the literature.
We have color-coded the points in this diagram to show that younger clusters ($\tau < 10^7$~yr) have been used to estimate $\Gamma$ in galaxies with high $\Sigma_{SFR}$ (blue dots) and older clusters ($\tau > 10^7$~yr) in galaxies with low $\Sigma_{SFR}$ (green dots).
This is a consequence of inadvertant biases in the selection of galaxies for such studies.
To date, $\Gamma$ has not been estimated from younger clusters in galaxies
with low $\Sigma_{SFR}$ or from older clusters in galaxies with high $\Sigma_{SFR}$.
As we show here, this selection bias, together with the decline of $\Gamma$ with age caused by the progressive disruption of clusters, has introduced a spurious correlation between $\Gamma$ and $\Sigma_{SFR}$.
%This is a serious problem because the age distributions of clusters decline with age, as a consequence of their progressive disruption.
%Thus, there are more stars in young clusters than in old clusters simply because there are more young clusters than old clusters.

The main purpose of this paper is to derive $\Gamma$ values by a homogeneous procedure for the same 8 galaxies for which we have previously derived CMF$/$SFR statistics.
This will allow us to test the claims that $\Gamma$ depends on properties of the galaxies such as $\Sigma_{SFR}$ and $\Sigma_{H_2}$. 
Part of our motivation comes from the fact that the existing results for $\Gamma$ are based on a heterogeneous set of assumptions and procedures, including the different age ranges for clusters in different galaxies discussed above.
In this paper, we adopt a homogeneous set of assumptions and procedures, including common age ranges for clusters, in order to avoid the biases that have crept into previous studies of $\Gamma$.

The remainder of this paper is arranged as follows.
In Section~2, we derive relations between the quantities $\Gamma$,
CMF$/$SFR, and the mass and age distributions, $\psi(M)$ and $\chi(\tau)$, of the clusters.
In Section~3, we determine new values of SFR, $\Sigma_{SFR}$, and $\Sigma_{H_2}$ for our sample galaxies,
and in Section~4 we summarize the cluster catalogs and present the age and mass distributions of the clusters.
In Section~5, we determine new values of $\Gamma$, 
%in the eight galaxies for which we have previously studied the CMF$/$SFR distributions. In Section~4 we 
and assess whether they or our previously derived CMF$/$SFR statistics show trends with any galaxy property.
%in $\Gamma_F$ and in the residuals of the CMF$/$SFR amplitudes with 
%star-forming properties of our sample galaxies, in particular 
In Section~6, we compare our new results with those from previous observational and theoretical studies,
and in Section~7 we discuss the physical implications of our results. 
In Section~8 we summarize our main conclusions.

\section{Relations Between Statistical Properties of Cluster Populations}

As explained in the Introduction, this paper is concerned with several statistical properties of cluster populations: the mass and age distributions, $\psi(M)$ and $\chi(\tau)$, the CMF normalized by the SFR (CMF$/$SFR), and the fraction of stars in clusters $\Gamma$.
Because clusters are progressively destroyed, $\Gamma$ will depend on the age interval over which it is determined, as shown in Figure~1.
Thus, it is essential to distinguish between the fraction of stars in {\em forming} clusters, which we henceforth denote by $\Gamma_F$, and the fraction of stars in {\em surviving} clusters, which we denote by $\Gamma_S$.
In this section, we derive some useful formulae for estimating $\Gamma_F$ and $\Gamma_S$ from observations.
We also show how these quantities are mathematically related to $\psi(M)$, $\chi(\tau)$, and CMF$/$SFR.
In the following sections, we present our observational determinations of $\Gamma_F$, $\Gamma_S$, and CMF$/$SFR, and we show that they have approximately the expected behavior.

With these issues in mind, we distinguish between two versions of the joint mass-age distribution of a cluster population: one for {\em forming} clusters, $f(M,\tau)$, and the other for {\em surviving} clusters, $g(M,\tau)$.
In general, $g(M,\tau)$ will be less than $f(M,\tau)$, except near $\tau=0$, because clusters are progressively destroyed by a variety of internal and external dynamical processes.
Both $f(M,\tau)$ and $g(M,\tau)$ are of theoretical interest, but only $g(M,\tau)$ is directly observable (since $f(M,\tau)$ includes all clusters that form, whether or not they survive to an age $\tau$).
%with current facilities 
%(the resolution and infrared sensitivity of NASA's upcoming JWST mission will allow studies of the youngest, embedded cluster phase in nearby galaxies).
All the statistical properties of cluster populations discussed in this paper can be derived from $g(M,\tau)$.
The mass function $\psi(M)$ is the integral of $g(M,\tau)$ over all $\tau$, while the age distribution $\chi(\tau)$ is the integral of $g(M,\tau)$ over all $M$.

The fraction of stars that form in clusters is simply the ratio of the formation rates of clusters and stars: $\Gamma_F = \mbox{CFR}/\mbox{SFR}$.
Another way to express this is in terms of the masses of recently formed clusters and stars in a small but common age interval $0 < \tau < \tau_{\alpha}$ that we will specify later:

\begin{equation}
M_S ( < \tau_{\alpha})~ =~ \tau_{\alpha}~ \mbox{SFR},
\end{equation}

\begin{equation}
M_C ( < \tau_{\alpha})~ =~ \int_{0}^{\tau_{\alpha}} \int_{0}^{\infty} M f(M,\tau)dM d\tau.
\end{equation}

\noindent Thus, we have

%\begin{equation}\label{eq:3a}
%\setcounter{equation}{3.1}
\centerline{~~~~~~~~~~~~~~~~~~~~~~~~~~~~~~~~~~$\Gamma_F = M_C(< \tau_{\alpha}) / M_S(< \tau_{\alpha}),~~~~~~~~~~~~~~~~~~~~~~~~~~~~~~~~~\mbox{(3a)}$}
%\end{equation}

\bigskip
%\begin{equation}\label{eq:3b}
%\setcounter{equation}{3.2}
\centerline{~~~~~~~~~~~~~~~~~~~~~~~~~~~~~~~~$= {\frac{1}{\tau_{\alpha}\mbox{SFR}} \int_{0}^{\tau_{\alpha}} \int_{0}^{\infty}} M f(M,\tau)dM d\tau.~~~~~~~~~~~~~~~~~~~~~~~~~~\mbox{(3b)}$}
%\end{equation}

\noindent If $\tau_{\alpha}$ is chosen to be small enough that disruption can be neglected, i.e., $f(M,\tau) \approx g(M,\tau)$ for $\tau < \tau_{\alpha}$, we will then also have

\begin{equation}
\setcounter{equation}{4}
\Gamma_F \approx  \frac{1}{\tau_{\alpha}\mbox{SFR}} \int_{0}^{\tau_{\alpha}} \int_{0}^{\infty} M g(M,\tau)dM d\tau.
\end{equation}

At this stage, it is important to note that $f(M,\tau)$, $g(M,\tau)$, and hence $\Gamma_F$ pertain to {\em all} compact clusters, irrespective of whether they are gravitationally bound or unbound.
Many compact clusters will either be born unbound or will become unbound by internal stellar feedback soon after they are born (often called ``infant mortality'').
The internal dynamical or crossing times of clusters vary widely, but $\tau_c \sim 10^6$~yr may be taken as a typical value.
N-body simulations show that once a cluster becomes unbound, it takes a time $\tau_d \sim 10 \tau_c \sim 10^7$~yr or more to dissolve into the surrounding stellar field (e.g., Baumgardt \& Kroupa 2007).
For much of this time, it will retain the appearance of a bound cluster, will be counted in cluster samples, and hence will be included in determinations of $g(M,\tau)$.

The choice of the age $\tau_{\alpha}$ in equations (3) and (4) involves several competing constraints.
On the one hand, $\tau_{\alpha}$ must be small enough that the dissolution of clusters can be neglected, as we have already noted.
It must also be small enough that temporal variations in the CFR and SFR can be safely neglected.
On the other hand, $\tau_{\alpha}$ must be large enough that the age interval $0 < \tau < \tau_{\alpha}$ includes enough clusters in real samples for accurate determinations of $\Gamma_F$.
%In practice, the value $\tau_{\alpha} \sim 10^7$~yr does a reasonable job of satisfying these constraints.
These constraints lead to $\tau_{\alpha} \sim 10^7$~yr.
This value of $\tau_{\alpha}$ also corresponds approximately to the period over which a new generation of stars produces ionizing radiation and hence H$\alpha$ emission.
% (thus the subscript $\alpha$).

We next consider some practical issues in determining $\Gamma_F$.
For a complete sample of clusters, with individual masses and ages $M_i$ and $\tau_i$, equation~(4) can be replaced by a discrete sum

\begin{equation}
\Gamma_F \approx \frac{1}{\tau_{\alpha} \mbox{SFR}} \sum_{i} M_i~~~~~~~~~~~ \mbox{for}~ \tau_i < \tau_{\alpha}.
\end{equation}

\noindent Most cluster samples are complete only above a limiting mass $M_{lim}$ set by the flux-detection limit and the distance of the galaxy.
For such mass-limited samples, the sum in equation~(5) must be broken into two parts: a sum over detected clusters with $M_i > M_{lim}$ and $\tau_i < \tau_{\alpha}$, and an integral over the mass function of undetected, less massive clusters.
Thus,

\begin{equation}
\Gamma_F \approx \frac{1}{\tau_{\alpha} \mbox{SFR}} {\bf \bigg[} \sum_{i} M_i + \int_{M_{min}}^{M_{lim}} M \psi(M)dM {\bf \bigg]}.
\end{equation}

\noindent The lower limit on the integral above represents the transition between clusters and individual stars, for which we adopt $M_{min}=10^2~M_{\odot}$.

The mass functions of young star clusters can be represented by a power law, $\psi(M) = A M^{\beta}$, over a large range of mass, $10^2~M_{\odot} \lea M \lea 10^6~M_{\odot}$, with nearly the same exponent $\beta \approx -2$ in different galaxies (Fall \& Chandar 2012).
Thus, equation~(6) becomes

\bigskip

%\begin{equation}\label{eq:7}
\centerline{~~~~~~~~~~~~~~~~~~
${\Gamma_F \approx \frac{1}{\tau_{\alpha} \mbox{SFR}} {\bf \bigg[} \sum_{i} M_i + \frac{A}{2+\beta}(M_{lim}^{2+\beta} - M_{min}^{2+\beta}) {\bf \bigg ]}}~~~ \mbox{for}~~ \beta \neq -2,~~~~~~~~~\mbox{(7a)}$}
%\end{equation}

\bigskip

%\begin{equation}\label{eq:7}
\centerline{~~~~~~~~~~~~~~~~~~~~~~~
${\approx \frac{1}{\tau_{\alpha} \mbox{SFR}} {\bf \bigg[} \sum_{i} M_i + A~ \mbox{ln}(\frac{M_{lim}}{M_{min}}) {\bf \bigg]}}~~~~ \mbox{for}~ \beta= -2.~~~~~~~~~~~~~~~~~~~\mbox{(7b)}$}
%\end{equation}

\noindent This is the equation we use to compute $\Gamma_F$.
We determine the normalization $A$ for each galaxy from the clusters more massive than $M_{lim}$.
For distant galaxies, for which $M_{lim}$ is large, the second term in the brackets of equation~(7) can be comparable to or even larger than the first term. We discuss corrections for dust attentuation to estimates of the SFR in Section~3.1 and to counts of clusters in Section~5.1.

We now consider the fraction of stars in surviving clusters $\Gamma_S(\tau_1,\tau_2)$ in an interval of age $\tau_1 < \tau < \tau_2$.
By analogy with equations (3) and (4), this is given by

\begin{equation}\label{eq:8}
\setcounter{equation}{8}
\Gamma_S(\tau_1,\tau_2) =  \frac{1}{(\tau_2-\tau_1)\mbox{SFR}} \int_{\tau_1}^{\tau_2} \int_{0}^{\infty} M g(M,\tau)dM d\tau.
\end{equation}

\noindent Because $g(M,\tau)dM d\tau$ is defined to be the number of surviving clusters in $(M,M+dM)$ and $(\tau,\tau+d\tau)$, equation~(8) is exact for all values of $\tau_1$ and $\tau_2$.
As we have already noted, in the special case $\tau_1=0$ and $\tau_2=\tau_{\alpha}$, the fractions of stars in forming and surviving clusters are nearly equal: $\Gamma_F \approx \Gamma_S(0,\tau_{\alpha})$.
When determining $\Gamma_S(\tau_1,\tau_2)$ from a mass-limited sample of clusters, the integrals in equation~(8) must be replaced by a sum over the detected clusters with $M_i > M_{lim}$ and $\tau_1 < \tau_i < \tau_2$ and integrals over the mass-age distribution of undetected, less massive clusters by straightforward extensions of equations~(6) and (7).

It is instructive at this stage to re-express $\Gamma_S(\tau_1,\tau_2)$ in an alternative but equivalent form.
With this in mind, we define the ``average'' mass function $\overline{\psi}$ 
of clusters with ages in the interval $\tau_1 < \tau < \tau_2$ as follows:

\begin{equation}
\overline{\psi}(M | \tau_1, \tau_2) = \frac{1}{(\tau_2-\tau_1)} \int_{\tau_1}^{\tau_2} g(M,\tau) d\tau.
\end{equation}

\noindent Multiplying this by $M$, integrating over all $M$, and using equation~(8), we obtain

\begin{equation}
\Gamma_S(\tau_1,\tau_2) = \frac{1}{\mbox{SFR}} \int_{0}^{\infty} M\overline{\psi}(M | \tau_1, \tau_2) dM.
\end{equation}

\noindent In a previous study, we named the function $\overline{\psi}(M | \tau_1, \tau_2)/\mbox{SFR}$ the CMF$/$SFR statistic (CFW15).
We now see from equation~(10) that $\Gamma_S(\tau_1,\tau_2)$ is simply the integral of $M\times$(CMF$/$SFR) over all $M$.
In our previous work, we found that CMF$/$SFR is nearly the same for the 8 galaxies we analyze in the present work (CFW15, Mulia et al. 2016).
Thus, we expect to find that $\Gamma_S(\tau_1,\tau_2)$ is also similar for these 8 galaxies.

The fractions of stars in forming and surviving clusters, $\Gamma_F$ and $\Gamma_S$, are closely related to the age distribution $\chi$, as we now demonstrate explicitly.
In previous work, we found empirically that the mass and age distributions, $\psi$ and $\chi$, are effectively independent of each other (Fall \& Chandar 2012).
Thus, the joint mass-age distribution of surviving clusters is, to a good approximation, separable:

\begin{equation}
g(M,\tau) = \psi(M) \chi(\tau).
\end{equation}

\noindent It helps at this stage to introduce the average value of $\chi$ over an age interval $\tau_1 < \tau < \tau_2$:

\begin{equation}
\overline{\chi}(\tau_1,\tau_2) = \frac{1}{(\tau_2 -\tau_1)}\int_{\tau_1}^{\tau_2} \chi(\tau) d\tau. 
\end{equation}

\noindent Inserting $g(M,\tau)$ from equation~(11) into equations~(4) and (8) and then using equation~(12) we derive the basic relation between $\Gamma_F$, $\Gamma_S$, and $\chi$:

\begin{equation}
\frac{\Gamma_F}{\Gamma_S(\tau_1,\tau_2)} = \frac{\overline{\chi}(0,\tau_{\alpha})}{\overline{\chi}(\tau_1,\tau_2)}.
\end{equation}

\noindent This result has a pleasing simplicity: apart from normalization factors, $\Gamma_S(\tau_1,\tau_2)$ and $\overline{\chi}(\tau_1,\tau_2)$ are the same function of $\tau_1$ and $\tau_2$.
Thus, if $\chi(\tau)$ declines with increasing $\tau$, as expected from the disruption of clusters, $\overline{\chi}(\tau_1,\tau_2)$ and hence $\Gamma_S(\tau_1,\tau_2)$ will decline with increasing $\tau_1$ and $\tau_2$.

To estimate $\Gamma_F/\Gamma_S(\tau_1,\tau_2)$, we proceed as follows.
In previous work, we found that the age distribution can be represented by a power law, $\chi(\tau) \propto \tau^{\gamma}$, over the age range $10^7 \lea \tau \lea 10^9$~yr with similar (but not exactly the same) exponents in different galaxies: $-1.0 \lea \gamma \lea -0.5$ (Fall \& Chandar 2012).
For ages below $\tau_{\alpha}\sim10^7$~yr, the shape of $\chi(\tau)$ is less certain; it could continue to rise toward $\tau=0$, or it could flatten off.
We recall that this age range includes both bound clusters and unbound clusters that have not yet dissolved away.
With this in mind, we consider two simple models for $\chi(\tau)$.
The first (model~1) is a pure power law with an exponent $\gamma > -1$ for all $\tau$, while the second (model~2) is flat for $0 < \tau < \tau_{\alpha}$ and a power law with an (unrestricted) exponent $\gamma$ for $\tau \geq \tau_{\alpha}$.
From equations~(12) and (13), we have

%\begin{equation}\label{eq:14}
model~1~~~~~~~~~~~~~~~~~~~~~~~~~
${\frac{\Gamma_F}{\Gamma_S(\tau_1,\tau_2)} = \frac{\tau_{\alpha}^{\gamma}(\tau_2-\tau_1)}{\tau_{2}^{1+\gamma} - \tau_{1}^{1+\gamma}}},~~~~~~~~~~~~~~~~~~~~~~~~~\mbox{(14a)}$
%\end{equation}

\bigskip

%\begin{equation}\label{eq:7}
model~2~~~~~~~~~~~~~~~~~~~~
${\frac{\Gamma_F}{\Gamma_S(\tau_1,\tau_2)} = (1+\gamma)\frac{\tau_{\alpha}^{\gamma}(\tau_2-\tau_1)}{\tau_{2}^{1+\gamma} - \tau_{1}^{1+\gamma}}}.~~~~~~~~~~~~~~~~~~~~~~~\mbox{(14b)}$

%\end{equation}

%\centerline{~~~~~~~~~~~~~~~~~~~~~~~~~~~~~~~~~~~~~~~~~~~~~~$
%\mathlarger{\frac{\Gamma_F}{\Gamma_S(\tau_1,\tau_2)} = \frac{(\tau_2-\tau_1)\tau_{\alpha}^{\gamma}}{\tau_{2}^{1+\gamma} - \tau_{1}^{1+\gamma}}}~~~~~~~~~~~~~~~~~~~~~~~~~~~~~~~~~~~\mbox{(13a)}$}

%\begin{equation}
%\frac{\Gamma_F}{\Gamma_S(\tau_1,\tau_2)} = \frac{(\tau_2-\tau_1)\int_{0}^{\tau_{\alpha}} \chi(\tau) d\tau}{\tau_{\alpha} \int_{\tau_1}^{\tau_2} \chi(\tau) d\tau}.
%\end{equation}

%\bigskip

%\begin{equation}
%%\setcounter{equation}{13}
%%\centerline{~~~~~~~~~~~~~~~~~~~~~~~~~~~~~~~~~~$
%\approx (\tau_{\alpha}/\tau_2)^{\gamma}~~~~~~ \mbox{for}~~ \tau_2 >> %\tau_1~~~~~~~~~~~~(13b)
%\end{equation}

%\centerline{~~~~~~~~~~~~~~~~~~~~~~~~~~~~~~~~~~~~~~~~$
%\approx (\tau_{\alpha}/\tau_2)^{\gamma}~~~~~~ \mbox{for}~~ \tau_2 >> \tau_1.~~~~~~~~~~~~~~~~~~~~~~~~~~~~~~~~~~~\mbox{(13b)}$}

\noindent These formulae, which differ by a factor $1 + \gamma$, are expected to bracket the true values of $\Gamma_F/\Gamma_S(\tau_1,\tau_2)$.

We now compare the predictions above with the observations plotted in Figure~1.
We identify $\Gamma_F$ and $\Gamma_S(10,100)$ with the observed $\Gamma$ for $0 < \tau < 10^7$~yr (blue dots) and $10^7 < \tau < 10^8$~yr (green dots), respectively.
Thus, we set $\tau_{\alpha}=\tau_1=10^7$~yr and $\tau_2=10^8$~yr in equations~(14a) and (14b).
For $\gamma=-0.7$, a typical exponent of the age distribution, these equations predict $\Gamma_F/\Gamma_S(10,100)\approx9$ (model~1) and $\Gamma_F/\Gamma_S(10,100)\approx3$ (model~2).
Smaller (more negative) values of $\gamma$ increase $\Gamma_F/\Gamma_S(\tau_1,\tau_2)$ and vice versa for both models.
For comparison, the median observed $\Gamma_F$ and $\Gamma_S(10,100)$ in Figure~1 are 23\% and 4.2\%, respectively; hence $\Gamma_F/\Gamma_S(10,100)\approx5$, well within the range spanned by models~1 and 2.
We note that the observed $\Gamma$ values plotted in Figure~1 have substantial uncertainties (because they were derived by heterogeneous procedures) and that the exponent $\gamma$ is not known in most cases (and may vary somewhat among galaxies).
Nevertheless, the rough agreement we find between the predicted and observed $\Gamma_F/\Gamma_S(10,100)$ already indicates that much of the claimed correlation between $\Gamma$ and $\Sigma_{SFR}$ is spurious.
We will make more detailed comparisons based on our homogeneously derived values of $\Gamma_F$, $\Gamma_S$, and $\gamma$ later in this paper.

\section{Galaxy Properties}

%\subsection{Star Formation Rates}

The galaxies in our sample are: LMC, SMC, NGC~4214, NGC~4449, M83, M51, Antennae, and NGC~3256, 
which includes irregulars, spirals, and mergers that span wide ranges in distance, luminosity or mass, SFR, and $\Sigma_{SFR}$.  
%The distances of these galaxies are presented in Table~1.
This sample, while small, is reasonably representative of nearby star-forming galaxies in general.  
Some of the basic properties of our sample galaxies, discussed in this section, are listed in Table~1.

\subsection{Star Formation Rates}

We determined the current SFR for each galaxy 
%within the area covered by each cluster catalog, 
from extinction-corrected H$\alpha$ luminosities (CFW15; Mulia et al. 2016).
H$\alpha$-based SFRs are sensitive to the most recently formed massive stars, with $\approx$90\% of the emission coming from stars younger than $\approx7$~Myr (e.g., Kennicutt \& Evans 2012).
%In Chandar et al. (2015) we determined the total SFR for each galaxy (except NGC~3256, which was presented in Mulia et al. 2016).
We use the most recent H$\alpha$ flux measurements listed in NED and correct for contamination by [NII] emission, and assume the distances given in column~2 of Table~1 when converting to H$\alpha$ luminosities.
In order to correct for attenuation by dust, we use the 
24$\mu$m flux, and the formula given in Kennicutt \& Evans (2012).
%\begin{equation}
%  L(H\alpha)_{\mbox{corr}} = L(H\alpha)_{\mbox{obs}} + 0.020 L(25\mu m)
%\end{equation}
%\noindent We then adopt their Equation~(12) and Table~1 to determine the star formation rate:
%\begin{equation}
%  \mbox{log~SFR} (M_{\odot}/\mbox{yr}) = \mbox{log}L(H\alpha)_{\mbox{corr}} - 41.27.
%\end{equation}
Their calibration is based on the STARBURST99 models (Leitherer et al. 1999), which assume solar metallicity, a Kroupa (2001) IMF, similar to the Chabrier IMF assumed for the clusters, and a constant rate of star formation. 
%The H$\alpha$-based SFRs are compiled in Tables~1 and 2.
 The resulting SFRs are listed in column~3 of Table~1 and column~2 of Table~2.
These cover a range of approximately three orders of magnitude, from $0.06~M_{\odot}$~yr$^{-1}$ for the SMC, to $50~M_{\odot}$yr$^{-1}$ for the merging NGC~3256 system.

In CFW15, we discussed and quantified the sources of uncertainty in H$\alpha$-based SFRs, which include flux measurements, corrections made for attenuation by dust, 
%from the assumption of Case B recombination, 
temporal variations in the star formation rate, uncertainties in the calibration used to convert the measured fluxes to SFR, and the leakage of Lyman continuum photons from the parent galaxy.
We found that our SFR estimates agreed to within $\approx50$\% of previously published H$\alpha$-based rates, and adopt this as our uncertainty here.  

The cluster catalogs used in this work provide near-full coverage of the SMC, NGC~4214, NGC~4449, M51, the Antennae, and NGC~3256.
%For the other galaxies in our sample, 
We determined the fractional coverage of the optically luminous portion of each galaxy in CFW15 and Mulia et al. (2016), and give these values in Column~4 of Table~1.

Because our sample includes irregular, dwarf, and merging galaxies, it is possible that the SFR over the last $\approx 400$~Myr may have been different from the current one determined from the H$\alpha$ luminosity.
If this is the case, our estimates in Section~5 for the fraction of stars in surviving clusters, $\Gamma_S(10,100)$ and $\Gamma_S(100,400)$, would be affected (but not those for $\Gamma_F$).
Therefore, we have also estimated the SFR for each galaxy from flux measurements at other wavelengths.
Far ultraviolet and infrared-based SFRs are sensitive to stars formed over a longer timescale than H$\alpha$-based determinations, approximately the last $\sim 100$~Myr and few$\times100$~Myr instead of just the last $\sim 10$~Myr.
We use the most recent GALEX far-ultraviolet and Spitzer 24$\mu$m fluxes from the literature,
and the formulae given in Kennicutt \& Evans (2012) to derive an extinction-corrected FUV-based SFR (by including the 24$\mu$m flux to correct for attenuation by dust) and the 24$\mu$m flux measurements alone to determine an infrared-based SFR.  
These are listed in Columns~3 and 4 of Table~2.
A comparison between different measurements for the same galaxy and waveband suggests that the fluxes are uncertain at the $\approx25$\% level.

The four closest galaxies in our sample, the LMC and SMC, NGC~4214, and NGC~4449, also have previously published CMD-based star formation histories.  We determine the average SFR between $10-100$~Myr ago and $100-400$~Myr ago from these analyses, and also compile these values in Table~2.
%In Figure~3, we show the SFRs in the different age intervals compiled in Table~2.
Table~2 shows some very interesting results.
If we compare the H$\alpha$-based ($\tau \lea 10$~Myr) SFR with the average $10-100$~Myr and $100-400$~Myr CMD-based ones, the LMC, SMC, NGC~4214, and NGC~4449 do not appear to have had strong variations over the past $\approx400$~Myr, with most of the values within 50\% of one another.
FUV and $24\mu$m-based SFRs, which we determined for NGC~4214 and NGC~4449, give similar results to those from the other techniques and tracers.
%These similarities suggest that the  between the average CMD-based SFR estimates and the FUV and $24\mu$m ones 

Figure~2 plots the SFRs listed in Table~2 in the three age intervals of interest: $\lea 10$~Myr (H$\alpha$-based SFR), $10-100$~Myr (CMD-based SFR for the four closest galaxies, FUV-based SFR for the others), and $100-400$~Myr (CMD-based SFR for the 4 closest galaxies, IR-based SFR for the others).
Interestingly, the SFRs do not vary strongly over this age range in any of our sample galaxies.
More importantly, there are no systematic trends with age; the SFR increases slightly with age in some galaxies and decreases slightly in others.
%Our finding is consistent with the conclusion reached by McQuinn et al. (2015)
%from an analysis of the CMDs of nearly 20 nearby dwarf starburst galaxies, that 'bursts' of star formation in starburst galaxies extend over longer $\approx500$~Myr timescales.
Our main conclusion from Figure~2 is that all of the galaxies in our sample appear to have had fairly constant {\em average} rates of star formation, to within $\approx50$\%, over the age intervals of interest here.
%, over the last $\approx400$~Myr.

%\subsection{Star Formation Rate Densities}
\subsection{Star Formation Rate Densities}

%In most previous works (both observational and theoretical), $\Gamma_F$ has been compared with the star formation rate per unit area, or $\Sigma_{SFR}$,
%measured in $M_{\odot}$~kpc$^{-2}$.
The rate of star formation per unit area, $\Sigma_{SFR}$ has been suggested to play an important role in the fraction of stars that form in bound clusters (e.g., Kruijssen 2012).
We have already summarized our method for determining the SFR of each galaxy.
Next, we determine the area covered by young clusters in each galaxy, using the same images as those used to produce the catalogs.
We do not include outlying regions where no 
%young $\tau \lea 10$~Myr 
young clusters are detected, and estimate that the uncertainty in our star-forming areas are $\approx50$\%.
We record these areas in column~5 of Table~1.
$\Sigma_{SFR}$ is determined by dividing the total SFR (corrected by the fractional coverage) by the area,
%(column~3 $\times$ column~4 divided by column~5),
and is listed in column~6.
The $\Sigma_{SFR}$ values cover approximately three orders of magnitude, from $\approx 9.2\times10^{-4}~M_{\odot}~\mbox{yr}^{-1}~\mbox{kpc}^{-2}$ 
for the dwarf irregular SMC to $\approx 1.0~M_{\odot}~\mbox{yr}^{-1}~\mbox{kpc}^{-2}$ for the merger NGC~3256.
The galaxies in our sample have an approximately linear relationship between SFR and $\Sigma_{SFR}$.
% follow an approximately linear relationship for the galaxies in our sample.

We also list previously published values for $\Sigma_{SFR}$ (from the compilation in Adamo et al. 2015) in column~7 of Table~1.
These are within a factor of two of our new determinations 
for the LMC, SMC, M83, M51, and NGC~3256; Adamo et al. (2015) did not list values for NGC~4449 or the Antennae.
%The previously published $\Sigma_{SFR}$ for M51 is a factor of 2 lower than our value, because the area used for the galaxy was very large, $\approx 580~\mbox{kpc}^2$.
There is no standard method for estimating the star-forming area within a galaxy, and if a significant area with no star formation is included, then $\Sigma_{SFR}$ can be underestimated. 
The previous estimate of $\Sigma_{SFR}$ for NGC~4214 is nearly six times lower than the one that we estimate here.  
This is at least in part due to the larger distance (and hence larger area) previously assumed for this galaxy.
Regardless, our conclusions do not change if the literature values for $\Sigma_{SFR}$ are used instead of our newly determined ones.

\subsection{Gas Densities}

The amount of molecular gas and how densely it is packed into galaxies might also play a role in star and cluster formation (e.g., Kruijssen 2012).
We list the average surface density of H$_2$ gas (in $M_{\odot}$~pc$^{-2}$) in Column~8 of Table~1, taken from the compilation in Kruijssen \& Bastian (2016) for seven of our galaxies, and from that in Kruijssen (2012) for NGC~3256.
These come from low-transition maps of CO, which are used to infer the distribution of H$_2$ molecular gas.
We update the values of $\Sigma_{H_2}$ for M51 (using data from Leroy et al. 2013) and the Antennae (using data from Zhang et al. 2001).
The values of $\Sigma_{H_2}$ extend from $\approx9~M_{\odot}$~pc$^{-2}$ in the SMC to $\approx 125~M_{\odot}$~pc$^{-2}$ in NGC~3256, a much smaller range ($\sim10\times$) than those of SFR and $\Sigma_{SFR}$ ($\sim10^3\times$).

%Despite the fact that our target galaxies cover three orders of magnitude in SFR and $\Sigma_{SFR}$, the range in gas density $\Sigma_{H_2}$ is significantly smaller, extending from 
%We find somewhat different values for M51 ($53~M_{\odot}~\mbox{pc}^{-2}$ 
%from Leroy et al. 2013, versus $30~M_{\odot}~\mbox{pc}^{-2}$ from Kennicutt 1998)
%and the Antennae (we calculate $125~M_{\odot}~\mbox{pc}^{-2}$ from the mass of molecular gas given in Zhang et al. (2001) versus the $76~M_{\odot}~\mbox{pc}^{-2}$ listed in Kruijssen \& Bastian), and list these additional estimates in parentheses.
%Our results, however, are not sensitive to the exact values of $\Sigma_{H_2}$ that we use for M51 and the Antennae.

\section{Star Cluster Catalogs and Properties}

\subsection{Observations}

The data and cluster selection criteria for our sample galaxies are presented and summarized in CFW15 and Mulia et al. (2016).
Candidate clusters were required to be compact but not to be round.
Here, we briefly summarize the catalogs used in this work, and refer the reader to the original work for more details.

{\it LMC and SMC:} We use the ground-based $UBVR$ catalogs from Hunter et al. (2003), where clusters were selected by visual examination of candidates compiled from previously published catalogs.
The catalogs include young clusters ($\tau \lea 10$~Myr) but not low density HII regions, with a total of 854 clusters in the LMC and 239 in the SMC.
In Chandar et al. (2010a), we estimated the age and mass of each cluster, and assessed the completeness of the catalog.
The data used by Hunter et al. (2003) covers $\approx70$\% and 90\% of the recent star formation in the LMC and SMC, as traced by H$\alpha$ emission (CFW15).

{\it NGC~4214:} We use a new catalog of 334 candidate star clusters, selected from images taken with $HST$/WFC3 as part of program GO-11360 which covers the galaxy in its entirety.
Cluster candidates were selected to be broader than the point-spread function (PSF), with close pairs of stars and background galaxies eliminated by visual inspection.
Aperture photometry in $UBVI$H$\alpha$ images was performed in a manner similar to that described in Chandar et al. (2010c) for cluster candidates in the nearby spiral galaxy M83.

{\it NGC~4449:} We use the $UBVI$H$\alpha$ catalog of 129 clusters published by Rangelov et al. (2011) 
based on $HST$ ACS/WFC ($BVI$H$\alpha$) and WFPC2 (U band) observations.  
The selection method was similar to that used for NGC~4214 and M83.  The observations cover nearly the entire galaxy.

{\it M83:} We use a new $UBVI$H$\alpha$ catalog of 3186 compact
clusters selected from seven fields observed with the $HST/WFC3$
(B. Whitmore et al., in prep). These fields cover $\approx60$\% of the
optically luminous portion of M83.  The cluster selection procedure
was similar to that followed for NGC~4214 and NGC~4449.

{\it M51:} We use the catalog of 3812 compact clusters published by Chandar et al. (2016), based on $HST$ ACS/WFC ($BVI$H$\alpha$) and WFPC2 (U band) observations. These cover $\approx90$\% of the optically luminous portion of M51.  

{\it Antennae:} We use the $UBVI$H$\alpha$ catalog published by Whitmore et al. (2010) based on $HST$ ACS/WFC ($BVI$H$\alpha$) and WFC3 ($U$ band) observations.  Clusters were selected to be objects brighter than $M_V=-9$, which eliminates nearly all individual, luminous stars.

{\it NGC~3256:} We use the $UBVI$H$\alpha$ catalog of 505 clusters published by Mulia et al. (2016) based on $HST$ observations that cover the entire main body of the galaxy.  Clusters were selected to be broader than the PSF.

Photometry for each cluster in each catalog was compared with predictions from stellar population models in the appropriate filters, in order to estimate the mass, age, and extinction of the cluster.\footnote{All galaxies in our sample have photometry in $UBVI$H$\alpha$, except for the LMC and SMC, which only had $UBVR$ photometry available.}
All of the cluster masses have been determined assuming a Chabrier (2003) IMF, including for M51 and the Antennae, for which the published masses assumed a Salpeter (1955) IMF.
%Details for each galaxy can be found in the references above.

\subsection{Age Distributions}

In Figure~3, we show the age distributions of mass-limited samples of star clusters for each galaxy studied here.\footnote{ All of the age distributions we have derived, in this and earlier papers, are for mass-limited subsamples of clusters.  
These can differ significantly from age distributions derived from luminosity-limited subsamples, as discussed in detail by Chandar et al. (2010b); see especially the discussion of the SMC in the Appendix of Chandar et al. 2010a).}
We have been careful to restrict the age range for each distribution to stay above the luminosity limit of the cluster catalog (which is set mostly by the distance to the galaxy).
For NGC~3256, we restrict our age-mass ranges to remain within the completeness limit for the dustier inner portion of the galaxy, as discussed in Mulia et al. (2016).

%We find that each age distribution follows a power law, $\chi(\tau) \propto dN/d\tau \propto \tau^{\gamma}$.
%These distributions decline steadily starting at very young ages, with no obvious dependence on the mass of the clusters, and all have exponents in the range $-1.0 \gea \gamma \gea -0.4$ (i.e., $\gamma \approx -0.7 \pm0.3$).
%We compile the best-fit value of $\gamma$ for each age distribution in Table~3.

Each age distribution can be represented by a simple power law, $\chi(\tau) \propto dN/d\tau \propto \tau^{\gamma}$, as found in previous studies (Fall \& Chandar 2012).
We list the best-fit exponents $\gamma$ in Table~3.
Most of them (but not all) are close to the median value $\gamma=-0.7$.
%Our best-fit values of $\gamma$ for M83 are very similar to those found by Silva-Villa et al. (2014) from an independent cluster catalog based on the same observations.
We emphasize that we have derived the exponents $\gamma$ of the age distribution self-consistently from the same data that we have used to derive the fraction of stars in forming and surviving clusters, $\Gamma_F$ and $\Gamma_S(\tau_1,\tau_2)$, and the normalized mass function CMF$/$SFR for the same galaxies (see Section~5).
For most galaxies, the power-law fits for each mass interval are nearly parallel, indicating that the mass and age distributions are nearly independent of each other.

The exponents we have derived for the age distribution in the 8 galaxies in our sample are generally in line with those derived in other studies;  most lie in the range $-1.0 \lea \gamma \lea -0.2$ (e.g., Fall et al. 2005; Whitmore et al. 2007; Bastian et al. 2012; Chandar et al. 2010a, 2014; Silva-Villa \& Larsen 2011; Messa et al. 2017).
The typical value for whole galaxies is $\gamma \approx -0.7\pm0.3$ (Fall \& Chandar 2012).
For M83, our galaxy-wide value of $\gamma\approx -0.4$ agrees nicely with that determined by Silva-Villa et al. (2014) from an independent cluster catalog based on the same observations.
For M51, our galaxy-wide value of $\gamma \approx -0.65$ is somewhat steeper than, but similar to, the $\gamma\approx -0.4$ recently found by Messa et al. (2017) from an independent cluster catalog based on different observations (with only $\sim60$\% of our coverage) and a different selection method.
There is some evidence for regional variations in $\gamma$.
In particular, the age distribution in M83 becomes flatter toward the outer parts of the galaxy (Silva-Villa et al. 2014).

The only exception to these results is the claim that the age distribution in M31 is essentially flat (Johnson et al. 2016).
We note, however, that the published plots actually show a gradually declining age distribution (see e.g., Figure~7 in Fouesneau et al. 2014, especially the bottom-left panel).
We have fitted a power law, $\chi(\tau) \propto \tau^{\gamma}$, to the M31 data in Johnson et al. (2016) over the age range $10~\mbox{Myr} < \tau < 250$~Myr, and find $\gamma \approx -0.25\pm0.1$ for mass ranges above $10^3~M_{\odot}$.
Thus, all galaxies studied so far have declining age distributions indicating the progressive disruption of their clusters.
We remind readers that even $\gamma=-0.2$ implies a disruption rate of $(1-10^{-0.2}) = 37$\% per decade (factor of 10) in age, while $\gamma=-0.4$, $-0.7$, and $-1.0$ imply disruption rates of 60\%, 80\%, and 90\%, respectively. 

%The exponents $\gamma$ we derive for the 8 galaxies in our sample are in line with the results derived by others, although larger variations in $\gamma$ are sometimes found on smaller spatial scales (e.g., Bastian et al. 2012; Adamo et al. 2015).
%In particular, our best-fit values of $\gamma$ for M83 are very similar to those found by Silva-Villa et al. (2014) from an independent cluster catalog based on the same observations.
%The main exception to this general result is for M31, for which Johnson et al. (2016) suggest $\chi(\tau) \approx \mbox{const}$, i.e. $\gamma\approx0$.
%However, when we fit over the age range $10~\mbox{Myr} < \tau < 250$~Myr, we find a declining age distribution with $\gamma \approx -0.25\pm0.1$ for different mass ranges above $10^3~M_{\odot}$.
%This value of $\gamma$ imples that nearly half the clusters are destroyed in a factor of 10 (decade) in age.

\subsection{Mass Functions}

The cluster mass functions are shown in Figure~4 in each of the three age intervals: $<10$~Myr, $10-100$~Myr, and $100-400$~Myr for all of our sample galaxies.
Each mass function can be represented by a simple power-law, $\psi(M) \propto dN/dM \propto M^{\beta}$, with no obvious curvature at either the high or low mass end.
The best-fit exponents are all close to the median value $\beta=-2.0$.
There is little variation in $\beta$ with age interval or from one galaxy to another, again indicating that the mass and age distributions are nearly independent of one another.
The lower limit of each mass function $M_{lim}$ is approximately set by the luminosity limit of the cluster catalog.
We have been careful to stay brighter than this limit for each age interval.
If incompleteness were a problem, we would expect the observed mass functions to flatten out near $M_{lim}$, contrary to what Figure~4 shows.
% and believe that incompleteness does not play a significant role in our mass functions, since none of the mass functions used here flatten at the low end. 
We compile values of $M_{lim}$ in Table~4.
The upper end of each mass distribution is not constrained.
%We find that the mass functions do not appear to depend on age, at least over the ranges plotted, for each galaxy.  The distributions all have exponents in the range $-2.2 \gea \beta \gea -1.8$ (i.e., $\beta \approx -2.0 \pm0.2$).
%We use these mass functions to determine $\Gamma$ in the next section.

\section{Correlations Between $\Gamma$ and CMF$/$SFR with Galaxy Properties}

An open question in this field is whether or not the fraction of stars forming in clusters varies with the global $\Sigma_{SFR}$, SFR, or $\Sigma_{H_2}$ of the host galaxy.
%conditions in the host galaxy, such as 
%, and if so, the strength of the correlation.
%Uncertainties in the various quantities clearly play a role in answering this question.
In this Section, we first use the cluster mass functions to determine the fraction of stars in clusters within the $<10$~Myr, $10-100$~Myr, and $100-400$~Myr intervals of age, and then determine whether these fractions vary with any of the galaxy properties discussed above.  We also compare with the residuals in the amplitudes of the CMF$/$SFR distributions, as a check on the results.

\subsection{The Fraction of Stars in Forming Clusters}

We now estimate the fraction of stars that form in clusters $\Gamma_F$ for each of the galaxies in our sample from equation~(7).
We recall that this is the ratio of the masses of newly formed clusters and stars, $\Gamma_F = M_C(<\tau_{\alpha}) / M_S(<\tau_{\alpha})$, in a small age interval $0 < \tau < \tau_{\alpha}$ [see equation~(3a)].
As we noted in Section~2, the choice of $\tau_{\alpha}$ is subject to several competing constraints, the most important of which is that all newly formed clusters, whether gravitationally bound or unbound, appear intact and are thus included in cluster catalogs.
These constraints lead to $\tau_{\alpha}\sim10^7$~yr.
In this case, $\Gamma_F$ is not sensitive to the exact choice of $\tau_{\alpha}$ because both $M_C(<\tau_{\alpha})$ and $M_S(<\tau_{\alpha})$ are affected in nearly the same way.
We adopt $\tau_{\alpha}=10^7$~yr.

Some of the youngest clusters, particularly those less than $1-3$~Myr old, are too deeply embedded in dusty natal material to be included in optical samples such as the ones we analyze here.
This suggests that the counts of optically detected clusters in the age range 
$0 < \tau < 10$~Myr should be corrected upwards by factors of $\sim10/9$ to $\sim10/7$ to account for the undetected clusters.
Corrections of this size are consistent with a multi-wavelength study that found a correspondence of $\sim85$\% between optically detected clusters and 6~cm radio sources in the Antennae galaxies (Whitmore \& Zhang 2002).
With this in mind, we adopt a conservative correction factor for missing clusters of $10/9$.
Others in this field have adopted correction factors up to $10/7$ (Goddard et al. 2010).
Evidently, this correction, which affects the absolute values of $\Gamma_F$, is uncertain at the factor of $1.15$ level.
We emphasize, however, that it has no effect on the {\em relative} values of $\Gamma_F$ among different galaxies, the main focus of this paper.
Our new determinations of $\Gamma_F$ are shown as the blue circles in Figure~5 and compiled in Table~5.

% ***LMC ****

The other main uncertainties in $\Gamma_F$, given the assumptions specified above, come from the observed CMFs and SFRs.
Uncertainties in the CMF for most galaxies are dominated by selection, 
particularly for $\tau \lea 10$~Myr clusters,
where catalogs from different groups are known to vary the most, due to crowding and other factors (e.g., Bastian et al. 2012; Chandar et al. 2014).
CFW15 quantified the uncertainties in the CMF at different ages, and found that the selection of clusters can affect the amplitude of observed mass functions in this age range at the $\approx10$\% level (but only at the $\approx4$\% level at older ages); we therefore assume that the estimates of the total mass in clusters are uncertain at this level, except for the LMC and SMC, the two galaxies with the lowest SFR and $\Sigma_{SFR}$ in our sample, where we include an additional 20\% uncertainty to account for stochastic sampling of the mass function based on Monte Carlo simulations.
CFW15 also included a detailed discussion and estimates of the different sources of uncertainty in the SFR determinations, which were summarized in their Table~2.
The SFR estimate for each galaxy was found to be uncertain by $\approx50$\%, based on a comparison with previous H$\alpha$-based determinations.
We adopt this level of uncertainty here when estimating the errors in $\Gamma$.
We do not, however, account for any systematic uncertainty in the SFR calibration, which may be overestimated for lower metallicity populations, 
because the ionizing luminosity increases by $\sim0.4\pm0.1$~dex for a tenfold decrease in the metallicity (see Kennicutt \& Evans 2012 \& references therein).
We recalculate $\Gamma_F$ using CMFs and SFRs that vary by the uncertainties discussed above, and take the minimum and maximum values as the most likely range of $\Gamma_F$.
Including all possible systematics and assumptions, we estimate that the final values of $\Gamma_F$ are uncertain by no more than a factor of $\sim1.7$.

\subsection{The Fraction of Stars in Surviving Clusters}

For the same galaxies, we also calculate and plot the fraction of stars found in surviving clusters, $\Gamma_S(10, 100)$ and $\Gamma_S(100,400)$.
For these calculations, we assume the same H$\alpha$-based SFRs as used for $\Gamma_F$, i.e. we assume a constant rate of star formation, since we found in the previous section that this is a reasonable assumption for the galaxies studied here.
We find mean values and standard deviations of 
$\Gamma_F\approx \Gamma_S(0,10)=24\pm9$\%,
$\Gamma_S(10,100)=4.6\pm2.5$\%, and 
$\Gamma_S(100,400)=2.4\pm1.1$\%.
The results for $\Gamma_S(10,100)$ and $\Gamma_S(100,400)$ do not change significantly if we adopt far-ultraviolet or infrared-based SFRs instead (see section~4.1.1).

In Figure~5, we see that the decline from 
$\Gamma_S(10,100)$ to $\Gamma_S(100,400)$ is smaller than the decline from 
$\Gamma_F=\Gamma_S(0,10)$ to $\Gamma_S(10,100)$.
Nevertheless, both declines are statistically significant because they occur for every galaxy in our sample (with the exception of NGC~4449)
from $\Gamma_S(10,100)$ to $\Gamma_S(100,400)$.
This decrease in the fraction of stars found in clusters is consistent with the observed decline in the cluster age distributions out to these ages.
The smaller decline from $\Gamma_S(10,100)$ to $\Gamma_S(100,400)$ than
from $\Gamma_S(0,10)$ to $\Gamma_S(10,100)$ reflects the differences in the age intervals, a factor of three in the first case and a factor of 9 in the second.

We now compare the predicted and observed ratios $\Gamma_F/\Gamma_S(10,100)$ and $\Gamma_F/\Gamma_S(100,400)$.
The median exponent of the age distribution from Table~3 is $\gamma=-0.7$.
As discussed in Section~2, for $\gamma=-0.7$, the predictions for $\Gamma_F/\Gamma_S(10,100)$ are $\approx9$ for model~1, and $\approx3$ for model~2.
The median values of $\Gamma_F\approx27$\% and $\Gamma_S(10,100)\approx5$\% for the 8 galaxies in our sample from Table~5 gives $\Gamma_F/\Gamma_S(10,100)\approx5$, comfortably within the range of predictions from models~1 and 2.
Interestingly, our results, which have been derived using a homogeneous procedure, are quite similar to the median $\Gamma$ values plotted in Figure~1, which were derived from heterogeneous procedures.
This suggests that the age interval used to determine $\Gamma$ has a stronger impact on the results than other assumptions (for example, the exact value of $M_{min}$, whether the assumed shape of the mass function is a power law or Schechter-function, etc.).
The predictions for $\Gamma_F/\Gamma_S(100,400)$ for $\gamma=-0.7$ are $\approx29$ for model~1 and $\approx8.5$ for model~2.
The ratio of the median $\Gamma_F$ to median $\Gamma_S$ for our galaxies is $\Gamma_F/\Gamma_S(100,400)\approx27/3\approx9$, again within the range spanned by the predictions from the models.
We also find rough agreement between the observed ratios $\Gamma_F/\Gamma_S(10,100)$ and $\Gamma_F/\Gamma_S(100,400)$ and the predicted ranges from models~1 and 2 for most individual galaxies in our sample, after taking into account the uncertainties on the exponent $\gamma$.

\subsection{Do $\Gamma_F$ or $\Gamma_S$ Correlate with Galaxy Properties ?}

Figure~5 shows some of the main results of this work.
The blue circles show our new determinations of $\Gamma_F$ plotted against $\Sigma_{SFR}$ of the host galaxy.
The best fit (shown as the gray line) is flat,
and indicates that {\em there is no significant trend in the fraction of stars in forming clusters across galaxies with increasing $\Sigma_{SFR}$.} 
The lower two gray lines which connect the green and red points show that there is no significant trend in the fraction of stars in {\em surviving} clusters across galaxies with increasing $\Sigma_{SFR}$.

The dotted lines connecting the blue, green, and red points for each galaxy highlight another important result:
{\em the fraction of stars in clusters depends strongly on the age interval that is used.}
The systematic decrease in this fraction with increasing age is a natural consequence of the declining age distributions observed for these galaxies,
which in turn reflects the disruption of clusters. 
These average observational results are similar to predictions from the simple disruption models discussed at the end of Section~2.
The labeled color bar on the left side of Figure~5 shows the good agreement between our observations and these predictions.

In the left panel of Figure~6, we show that our new results 
differ from the relationship found by Goddard et al. (2010) and plotted as the dashed line,
where $\Gamma$ increases approximately linearly
by a factor of $\approx10$ over the range covered by our data.
One key difference is that our work compares results made from the same age intervals in different galaxies.
Our study presents the first determination of $\Gamma$ made using clusters younger than 10~Myr in galaxies with low $\Sigma_{SFR} \lea 0.003$ (i.e., the LMC and SMC, since Goddard et al. 2010 published values for $\Gamma_S(10,100)$ rather than $\Gamma_S(0,10)$).
We further discuss the reasons for the discrepancy between our new results for $\Gamma$ and the previous ones in Section~6.

% Gamma vs. SFR
The middle panel of Figure~6 plots $\Gamma$ versus SFR.
%, since the SFR determinations have lower uncertainties than those for $\Sigma_{SFR}$.
%For this reason, Chandar et al. (2015) preferred the SFR over the $\Sigma_{SFR}$ when comparing the results for their results of the CMF$/$SFR statistic between galaxies.
%The middel panel of Figure~4 shows our new values of $\Gamma_F$ versus the SFR of the galaxies.
The blue points and best fit line again show no statistically significant correlation, indicating that $\Gamma_F$ does not increase systematically with the overall SFR (as determined from H$\alpha$) of the host galaxy for our sample.
%The Spearman correlation coefficients confirm this.
%This result is consistent with the absence of a correlation between $\Gamma_F$ and $\Sigma_{SFR}$, since there is an approximately linear relationship between $\Sigma_{SFR}$ and SFR for the galaxies in our sample.
% Gamma vs. gas density
The blue circles in the right panel of Figure~6 show that there is also no statistically significant trend between $\Gamma_F$ and $\Sigma_{H_2}$.
%$\Sigma_{H_2}$ is the preferred parameter in simulations, and $\Sigma_{SFR}$ is calculated from it by assuming the Schmidt-Kennicutt relation (e.g., Kruijssen 2012).
%While our sample galaxies span a range of $10^3$ in $\Sigma_{SFR}$ and SFR, they only span a range of $\approx10$ in gas density.
%In simulations, it is generally the density of the molecular gas that is the preferred parameter (e.g., Kruijssen 2012), with $\Sigma_{SFR}$ being determined by assuming the Schmidt-Kennicutt relation.
%Significantly smaller / lower range of gas densities when compared with star formation densities.
%The blue circles and associated fit in the right panel of Figure~3 
%show no statistically significant trend.
% between $\Gamma_F$ and $\Sigma_{H_2}$.
We conclude that our sample {\em does not reveal any trends in the fraction of stars that form in clusters (bound or unbound), with $\Sigma_{SFR}$, SFR, or $\Sigma_{H_2}$ in the host galaxy, over a fairly large range of these properties.} %found in star-forming galaxies out to $\approx 40$~Mpc.} 

Next, we compare the results from older surviving (and hence bound) clusters, $\Gamma_S(10,100)$ (green squares) and $\Gamma_S(100,400)$ (red triangles), which encode information about their {\em formation plus disruption} rates, with $\Sigma_{SFR}$ (left panel), SFR (middle), and $\Sigma_{H_2}$ (right).
%The green squares in each panel of Figure~3 show our new determinations for $\Gamma_{10-100}$, and the red triangles for $\Gamma_{100-400}$.
%The differences in the values calculated from different age intervals for the {\em same galaxy} are highlighted by the vertical dotted lines in the middle panel, where the blue circles ($\Gamma_F$, calculated from $\tau < 10$~Myr clusters) are significantly higher than the green squares ($\Gamma_S$ from $10 \leq < \tau < 100$~Myr clusters), which are higher than the red triangles ($\Gamma_S$ from $100 \leq \tau \leq 400$~Myr clusters).
%The best fits for $\Gamma_{10-100}$ and $\Gamma_{100-400}$ shown in Figure~3
%versus 
%$\Sigma_{SFR}$ (left panel), SFR (middle), and $\Sigma_{H_2}$ (right) of the host galaxy are also shown.
Our study presents the first determinations of $\Gamma_S(10,100)$ and $\Gamma_S(100,400)$ in galaxies with high $\Sigma_{SFR} \gea 0.03$ (i.e., the Antennae and NGC~3256), since Goddard et al. (2010) and Mulia et al. (2016) published values for $\Gamma_S(0,10)$ rather than $\Gamma_S(10,100)$ in NGC~3256. 

We find no statistically significant correlation for any of the fits in Figure~6. 
%{\bf This remains true if we adopt the $10-100$~Myr CMD-based or FUV-based SFR for the $\Gamma_S(10,100)$ calculations, and the $100-400$~Myr CMD-based or infrared-based SFR for the $\Gamma_S(100,400)$ ones.}
$\Gamma_S(10,100)$ shows a very weak ($\approx2\sigma$) trend with $\Sigma_{H_2}$, but not with $\Sigma_{SFR}$ or SFR. 
If instead we use 
%better match the 
CMD-based SFRs for the 4 closest galaxies and the FUV-based ones for the 4 more distant ones, even this weak trend disappears.
Values of $\Gamma_S(100,400)$ also show no statistically significant correlation with any galaxy property, regardless of whether we adopt the H$\alpha$-based or CMD-based SFRs for the 4 closest galaxies and infrared-based SFRs for the 4 more distant ones.
%Given the similarities in the SFR values found using different age tracers, we believe that variations in the star formation histories in these galaxies have a minimal impact on our results. }
Taken together, our results for $\Gamma_F$ and $\Gamma_S$ suggest that not only is there no systematic variation in the fraction of stars forming in clusters across galaxies with a large range in $\Sigma_{SFR}$, SFR, or $\Sigma_{H_2}$, {\em there is also no significant systematic variation in the fraction of stars in surviving clusters, even though these fractions are strongly affected by the disruption of the clusters.  }
These results are consistent with our proposal that both the formation and disruption of clusters are governed by quasi-universal processes
%The lack of correlations and the similarity in the values suggests that {\em the formation plus disruption rates of clusters in different galaxies are not strongly affected by the $\Sigma_{SFR}$, SFR, or the $\Sigma_{H_2}$ of the host galaxy.}
%Our data do not however, rule out weak correlations, and a larger galaxy sample will be needed to establish any trends.

%In summary, {\em we do not find evidence for statistically significant correlations between our new values of $\Gamma_F$ and $\Sigma_{SFR}$, SFR, or $\Sigma_{H_2}$, despite the fact that these galaxies cover a fairly large range in these properties.}
%Our experiments suggest that previous works that showed strong correlations were biased and showed systematic trends in
%$\Gamma_F$ with $\Sigma_{SFR}$ because different age ranges were used in different galaxies.

\subsection{Do Residuals in the CMF$/$SFR Amplitude Correlate with Galaxy Properties ?}

We recently developed a new method, which we call CMF$/$SFR, to compare the formation rates of stars and clusters (CFW15).
In this method, we compare the mass functions, $\psi(M) = dN/dM$, of recently formed clusters in different galaxies, before and after dividing by the star formation rates, and use this as a check on our new results for $\Gamma_F$.

The top-left panel of Figure~7 shows the observed CMFs for very young clusters ($\tau < 10$~Myr) for the eight galaxies considered in this work.
%For each mass function, an equal number of clusters from 3 to 7 are included in each bin, depending on the size of the cluster population.
%As a result, each data point has similar uncertainties, but different bin widths.
%The low end of each mass function is set by the luminosity limit of the cluster catalog (which is set mostly by the distance to the galaxy).
%We have been careful to stay brighter than this limit for each age interval and believe that incompleteness does not play a significant role in our mass functions or $\Gamma$ calculations (note that none of the mass functions flatten at the low end). 
%The upper end of each mass distribution is not constrained.
The amplitudes of the cluster mass functions reflect differences in the sizes of the cluster populations among the galaxies, and span a vertical range of $\approx10^3$.
The observed mass functions can be represented by featureless power laws, $dN/dM \propto M^{\beta}$, with $\beta \approx -1.9$, but each with a different normalization.
Some works have suggested that there is an exponential steepening at the upper end of the mass function for $M\gea M_C$, where $M_C$ varies between a few$\times10^4~M_{\odot}$ and $\approx10^6~M_{\odot}$ depending on the type of galaxy (e.g., Portegies Zwart, McMillan, \& Gieles 2010). 
However, these claims are based on the absence of just a handful of clusters relative to an extrapolated power law, and are not statistically significant.
%), and therefore do not affect our CMF$/$SFR derivations.
There is no evidence for a cutoff or steepening in the mass functions presented here.
The top-right panel in Figure~7 shows the CMF$/$SFR distributions with H$\alpha$-based SFRs.
These all lie very close to one another in the vertical direction.

We apply the technique developed in CFW15 to quantify the observed scatter, by fitting the CMF for each galaxy in the form:
%\begin{equation}
$dN/dM = A \times \mbox{SFR} \times (M/10^4~M_{\odot})^{-1.9}$.
%\end{equation}
These fits are shown as the solid lines in the panels on the right in Figure~7.
The coefficient $A$ measures the proportionality between the number of clusters and the SFRs.
The dispersion $\sigma(\mbox{log}~A)$ in the best fit values of log~$A$ quantifies the scatter in the amplitudes and hence in the CMF$/$SFR relation among the galaxies.
If the amplitudes of the cluster mass functions were exactly proportional to the star formation rates in their host galaxies, then the distributions in the top-right panel (for $\tau < 10$~Myr clusters) would all lie on top of one another, forming a single sequence.
This is very nearly what we find, with a scatter of only $\sigma(\mbox{log}~A)=0.23$, similar to the dispersion expected from errors in the CMFs and SFRs.

Although to first order the CMF$/$SFR distributions have very similar amplitudes, it is conceivable that there are weak correlations with properties of the host galaxy.
In Figure~8 we present the residuals in the amplitudes log~$A$ versus $\Sigma_{SFR}$, SFR, and $\Sigma_{H_2}$ of the host galaxy.
If the formation of the clusters were only proportional to the SFR of the host galaxy, then we would not expect to observe any trends in these diagrams.
If, on the other hand, the fraction of stars born in clusters were to increase with $\Sigma_{SFR}$, SFR, or $\Sigma_{H_2}$, this would manifest itself as a statistically significant increase in the residuals of log~$A$ with these parameters.
Figure~8 does not reveal any statistically significant trends in the residuals of the CMF$/$SFR amplitudes for $\tau < 10$~yr clusters (blue circles) with any of these galactic properties.
This is consistent with and supports our new results for $\Gamma_F$.

We can use older, surviving clusters to test the similarity of the {\em formation plus disruption} rates in our galaxies by comparing the functions 
$\overline{\psi}(M|\tau_1,\tau_2)/\mbox{SFR}$.
The middle and bottom-right panels in Figure~7 show that these distributions for $10-100$~Myr and $100-400$~Myr clusters also lie very close to one another, with dispersions $\sigma(\mbox{log~A})=$ 0.27 and 0.24, respectively, when we use the H$\alpha$-based SFRs.
The dispersions are similar if we use the same combinations of CMD, FUV and IR-based SFRs as in Section~5.3.
%This is the case, even though we have used the H$\alpha$-based SFRs, which could potentially introduce additional uncertainties for these older age intervals.

Figure~8 shows residuals in the $\overline{\psi}(M|\tau_1,\tau_2)/\mbox{SFR}$
amplitudes with galaxy properties as the green squares ($10-100$~Myr)
and red triangles ($100-400$~Myr).  This plot does not reveal any
statistically significant increasing trends in the residuals of the CMF$/$SFR
amplitudes for these older clusters with $\Sigma_{SFR}$, SFR, or
$\Sigma_{H_2}$.  In fact, the only weak ($<3\sigma$) trends (for
$\Sigma_{SFR}$ for $10-100$~Myr and $100-400$~Myr clusters), are decreasing.  The results are similar
  if we use the CMD, FUV, and IR-based SFRs.
% as in Section~4.2.
Overall, we find that these trends are consistent with the estimated observational uncertainties alone.\footnote{The results for NGC~3256 are the least certain,
  because it is the most distant galaxy in our sample.  However, we do
  not find any statistically significant correlations among any of the
  parameters plotted in Figure~8 if we remove NGC~3256 from our fits.}
We conclude that the CMF$/$SFR method gives results that are
consistent with those found for $\Gamma_F$ and $\Gamma_S$.

%\subsection{DISCUSSSION}

\section{Comparison With Previous Results}

\subsection{Observational Results}

In the previous section, we found that the two different (but related) methods, $\Gamma$ and CMF$/$SFR, give consistent results for the galaxies in our sample, and that neither method reveals a significant trend for the proportion of star formation in clusters with $\Sigma_{SFR}$, SFR, or $\Sigma_{H_2}$ among different galaxies.

Our new results for $\Gamma$ appear to contradict previously published ones, which have reported a strong correlation on galaxy scales between $\Gamma$ and $\Sigma_{SFR}$ and between $\Gamma$ and $\Sigma_{H_2}$ (e.g., Goddard et al. 2010; Kruijssen 2012; Adamo et al. 2015; Kruijssen \& Bastian 2016; Johnson et al. 2016).
However, we find that our results for $\Gamma$ actually agree fairly well with previously published values {\em when the same set of assumptions are used, in particular the same interval of age}. We also find that the specific catalog or method used to select the clusters does not have much affect on the results.
For the LMC and SMC, we find $\Gamma_S(10,100)$ values of $5$\% and 
$3$\%, respectively, from our mass and age estimates,
% when we use the parameters listed in Table~3, 
and 7\% and 5\% from the Hunter et al. (2003) mass and age estimates.
These are quite similar to the $\Gamma_S(10,100)$ values of $5.8\pm0.5$\% and $4.2^{0.2}_{0.3}$\% found for the LMC and SMC by Goddard et al. (2010) using the same cluster catalogs.
% (but much lower than the $\Gamma_F$ value of $\approx35$\% for the LMC that we discussed in Section~3).
For M83, we apply our method to the independent catalog published by Silva-Villa et al. (2014), and find $\Gamma_S(0,10)=17$\%, $\Gamma_S(10,100)=13$\%, and $\Gamma_S(100,400)=4$\%.
These are roughly similar to the values of 12\%, 10\%, and 2\% determined in Section~5 from our own M83 cluster catalog, and also to the value of $\Gamma_S(0,10)=18$\% determined by Adamo et al. (2015) from the Silva-Villa catalog with similar assumptions. 
This comparison of $\Gamma$ results for M83 between different works is particularly revealing because Silva-Villa et al. (2014) applied different criteria to select clusters, requiring that they be round in an attempt to select only bound clusters, while we did not.
%The results suggest that details of cluster selection may not play a significant role in $\Gamma_F$ determinations.
These results suggest that $\Gamma$ may not be particularly sensitive to the details of cluster selection.\footnote{The selection of clusters in only three out of the 19 previously studied galaxies compiled in Table~6 (M83, IC10, and NGC~2997) attempted to discern whether or not they were bound based solely on their appearance.}

Why have previous works (e.g., Goddard et al. 2010; Kruijssen 2012; Adamo et al. 2015) found a different result, one where $\Gamma$ varies strongly with $\Sigma_{SFR}$ and $\Sigma_{H_2}$?
While earlier studies also integrated over the CMF, as we have done here, they made a variety of different assumptions for different parameters when calculating $\Gamma$, 
most importantly the age interval (shown in Figure~1), the lower mass cutoff $M_{min}$, and the assumed shape of the cluster mass function (among others). 
In Table~6, we list previously published values for the fraction of stars found in clusters in 19 galaxies, taken from the recent compilation by Adamo et al. (2015), plus the newly published result for M31 from the PHAT team (Johnson et al. 2016).
We also list the assumptions used in each case, taken from the original work.
There is a large range, from just $1-3$\% for some galaxies all the way up to 50\% for others.

Figure~1 reveals a simple bias that has propagated through the literature, leading to an apparent strong trend between $\Gamma$ and $\Sigma_{SFR}$.
We now examine this diagram in more detail.
This figure plots previously published values,
as compiled in Table~6, but color-coded by the age interval that was used.
There is a striking trend: the $\Gamma_S(0,10) \approx \Gamma_F$ values (determined from very young $\tau < 10$~Myr clusters) shown in blue are systematically higher than the $\Gamma_S(10,100)$ values shown in green,
%(identical to the trend that we found in Section~3), 
and {\em previous works have preferentially determined $\Gamma_S(10,100)$ for galaxies with lower $\Sigma_{SFR}$ but $\Gamma_S(0,10)$ for galaxies with higher values of $\Sigma_{SFR}$, and then plotted $\Gamma_S(0,10)$ and $\Gamma_S(10,100)$ values all together.}
It is not difficult to understand why these choices might have been made.
Galaxies with high $\Sigma_{SFR}$ tend to be farther away, and hence samples are usually restricted to the brightest clusters, which tend to be young.
Meanwhile, galaxies with low $\Sigma_{SFR}$ tend to have relatively few very young clusters.
Unfortunately, these choices have led to systematic biases that are (mostly) responsible for the strong trend that has been claimed in many previous works.
%{\bf Another potential issue is that the galaxies with the highest values determined for $\Gamma_F$, ESO 338-IG04, SBS 0335-052E, and Haro~11, are not only some of the most distant galaxies with $\Gamma_F$ determinations, they also have the poorest quality observations, since they were observed with the WFPC2 camera rather than the more modern ACS or WFC3 on board the $HST$.
In contrast, by determining $\Gamma_S(0,10)$, $\Gamma_S(10,100)$, and $\Gamma_S(100,400)$ for all eight galaxies in our sample, we have been able to compare results determined from the same age range among the different galaxies.
% and have led to an incorrect picture for the relationship between star and cluster formation among different galaxies.

Table~6 demonstrates that, even beyond the choice of an appropriate age interval, a variety of different assumptions have been made for $\Gamma_F$ calculations in different galaxies.
%For example, some works assume a Kroupa IMF while others assume Salpeter. 
%The corresponding masses of the clusters and hence the $\Gamma_F$ values differ by $\approx40$\%.
%For the galaxies with the highest $\Sigma_{SFR}$ and previous estimates of $\Gamma_F$, ESO~338-IG04, SBS~0335-052E, and Haro~11, a Salpeter IMF was assumed.
%If a Kroupa or Chabrier IMF had been assumed instead, as we have done here, the $\Gamma_F$ values would be $\approx 38$\% (as initially published by Adamo et al. 2010), 
%quite similar to the $\Gamma\approx35$\%  that we find for the LMC, a galaxy with one of the lowest $\Sigma_{SFR}$ in our sample.
While most works have assumed a minimum cluster mass $M_{min}=100~M_{\odot}$, different values have been assumed for a few of the galaxies.
Some works have assumed that the cluster mass function is a pure power law, while others have assumed that it has a Schechter-like downturn at the high mass end, which can introduce differences in $\Gamma$ at the $\approx10$\% level, depending on the adopted $M_C$ value.
%of only three galaxies have 
In future studies, it is critical that authors compare $\Gamma$ values (and CMF$/$SFR statistic) among galaxies determined {\em from a consistent set of assumptions, as we have done here.}
We recommend that the interval $\tau \lea 10^7$~yr be used to calculate $\Gamma_F$ when one is interested in the fraction of stars {\em in forming} clusters (bound and unbound), but note that older intervals of age can be used to calculate $\Gamma_S$, the fraction of stars {\em in surviving} clusters at later times.
Regardless, the same age interval should be used to calculate the fraction of stars in clusters when comparing results among different galaxies. 

At this point, it is instructive to look in more detail at a specific galaxy with a well-studied cluster population.
%It is instructive to look in more detail at a specific galaxy with a well-studied cluster population.
The LMC has formed many massive, young compact clusters, including the well-known R136 in the 30 Doradus nebula ($M\sim10^5M_{\odot}$), and others such as H88-267, H88-301, KMK88-88, NGC~2080, BSDL2596, NGC~2074, and SL360
($M\sim10^4~M_{\odot}$).
Well over 100 other very young clusters are known throughout the LMC.
In Figure~9 we show a small, approximately $1\times2$~kpc portion of the LMC imaged in H$\alpha$, covering the Southern Molecular Ridge (e.g., Indebetouw et al. 2008).
R136 and several other clusters clearly have H$\alpha$ emission, as expected for clusters younger than $\tau \lea 10$~Myr, a good check on the age estimates for clusters in the LMC.
These young star-forming regions continue along the ridge to the south well beyond this image.
We find a total mass of $\approx 3.55\times10^5~M_{\odot}$ for 
clusters with estimated ages $\tau \lea 10$~Myr and masses down to $M_{lim}=10^3~M_{\odot}$, from our age-mass estimates and those determined independently by Hunter et al. (2003) for their LMC catalog.
The LMC has a current SFR of $0.25~M_{\odot}$~yr$^{-1}$ (see Table~1),
which implies a total stellar mass, $\tau_{\alpha}~\times \mbox{SFR}=2.25\times10^6~M_{\odot}$ formed over the last $\tau_{\alpha}=10$~Myr.
The observed clusters more massive than $M_{lim}=10^3~M_{\odot}$ contain $\approx16$\% of this mass
($3.55\times10^5/2.25\times10^6$).
The correction for clusters with masses between $M_{lim}$ and $M_{min}=10^2~M_{\odot}$ and for very young, highly obscured clusters ($\sim 10/9$; see the discussion in Section~5.1) brings the estimated fraction of stars formed in clusters to $\approx27$\%.
{\em Interestingly, the single most massive young cluster in the LMC, R136,  alone accounts for 
$\approx4$\% of the stellar mass formed in this period
($10^5/2.25\times10^6$).}

%In Section~3, we found a value of $\Gamma_F\approx35$\% for the LMC, a 
Our value of $27$\% is approximately 5 times higher than the one previously published by Goddard et al. (2010), because they calculated $\Gamma_S(10,100)$ rather than $\Gamma_F$.
Hence, using the same $\tau <10$~Myr interval for both the LMC and the galaxies with high $\Sigma_{SFR}$ and SFR gives essentially the same $\Gamma_F$ values: $\approx27$\% for the LMC and $\approx26$\% for the average of the three galaxies in our sample with the highest $\Sigma_{SFR}$.

%We find similar results for $\Gamma_F$ if we use the Hunter et al. (2003) age and mass estimates for clusters in the LMC instead of our own.  
%If we had assumed $\tau_{\alpha}=10$~Myr instead, then $\Gamma_F$ for the LMC would be reduced by $30$\% to $\Gamma\approx25$\%.

\subsection{Theoretical Predictions}

Our observational results present a challenge to theoretical models of the formation and early evolution of star clusters.
In particular, Kruijssen (2012) predicts that the fraction of stars that form in bound clusters has a strong dependence on $\Sigma_{SFR}$, shown by the dotted line in Figures~1 and 6.
Instead, we find that the fraction of stars in clusters of different ages, measured by $\Gamma_F$, $\Gamma_S(10,100)$, and $\Gamma_S(100,400)$, has essentially no dependence on their galaxy-wide stellar and interstellar environments, as measured by SFR, $\Sigma_{SFR}$, and $\Sigma_{H_2}$.

Based on $\Gamma_F$ alone, this test is inconclusive, because it includes an unknown mixture of bound and unbound clusters younger than $\sim10^7$~yr ($\sim10$ crossing times) (Kruijssen \& Bastian 2016).
Up to this age, both bound and unbound clusters remain relatively compact and virtually impossible to distinguish from each other, as demonstrated by N-body simulations (Baumgardt \& Kroupa 2007).
However, after $\sim10^7$~yr, only bound clusters remain compact, while unbound clusters expand and dissolve in the surrounding stellar field.
Thus, $\Gamma_S(10,100)$ and $\Gamma_S(100,400)$ represent the fraction of stars that form in bound clusters that survive to ages 10---100~Myr and 100---400~Myr ($\sim10-100$ and $100-400$ crossing times), respectively.
Since the survival probabilities of clusters (determined by the shapes of their age distributions) are similar among different galaxies, we conclude that the fraction of stars that form in bound clusters is also similar, in contradiction with the Kruijssen (2012) model.

On the other hand, the near constancy of $\Gamma_F$, $\Gamma_S(10,100)$, and $\Gamma_S(100,400)$ is fully consistent with, and even expected for, the quasi-universal model.
In this model, the formation and disruption of clusters depend mainly on local processes in the interstellar medium, and these are assumed to operate in much the same way from one galaxy to another.
Thus, the cluster formation rate is simply a constant $\Gamma_F$ times the star formation rate.
Similarly, the (fractional) disruption rate of clusters is also nearly constant, leading to the near constancy of $\Gamma_S(10,100)$ and $\Gamma_S(100,400)$.

\section{Discussion}
%\section{Discussion and Conclusions}

% both methods give consistent results: similarity star/cluster formation
We have used two different techniques, $\Gamma_F$ and CMF$/$SFR,
to study the relationship between star and cluster formation in eight galaxies with a large range of star-formation and gas properties.
We find that when applied consistently among galaxies, both techniques give similar results.

To first order, we find that star and cluster formation rates are proportional to one another, with a similar $\Gamma_F\approx24\pm9$\% fraction of stars forming in compact clusters in galaxies that cover a large range in $\Sigma_{SFR}$, SFR, and $\Sigma_{H_2}$.
Neither technique supports the strong correlations found in previous works,
which appear to have been the result of systematic biases.
However, with only eight galaxies in our sample, we cannot rule out that undetected weak trends may be present.  
A similar study, but with a larger sample of galaxies, is needed to detect and quantify any such trends.

We also find that the fraction of stars in surviving clusters, quantified here by $\Gamma_S(10,100)$ and $\Gamma_S(100,400)$, declines rapidly with age.
Based on the work and conclusions presented in Sections~5.3 and 5.4, this effect is much too strong to be explained by variations in the star formation rates in our galaxies, and instead must reflect the disruption of the clusters.
This conclusion is not surprising, given the declining shapes of the cluster age distributions shown in Figure~3.
%There are several indications that the SFRs have been fairly constant over the last few 100 Myr in these galaxies.
%{\bf First, the analysis of the colors and magnitudes of thousands to millions of individual stars in the LMC, SMC, NGC~4214, and NGC~4449 results in very similar average rates of star formation in the intervals $10-100$~Myr and $100-400$~Myr (compiled in Table~2).
%These values are quite similar to those determined from extinction-corrected H$\alpha$ luminosities, which gives the average star formation rate over the $1-10$~Myr interval.}
%while the SFHs in these galaxies have varied, this variation has been within a factor of $\approx$2 (Harris \& Zaritsky 2004, 2009).
%Second, some simulations of the merging Antennae galaxies suggest that the star formation history has been fairly flat for the last few$\times10^2$Myr (Karl et al. 2011).
%Third, the similarity between SFRs determined from H$\alpha$ and from far ultraviolet fluxes, which are sensitive to stars with ages $\tau < 10$~Myr and $\tau < 100$~Myr, respectively, suggests that the SFR has not varied dramatically over the last $\approx100$~Myr.
%This leaves disruption as the most likely explanation for the  observed drop from $\Gamma_F$ to $\Gamma_S(10,100)$ to $\Gamma_S(100,400)$. 

Johnson et al. (2016) found a modest increase in
$\Gamma_S(10,100)$ (from $\approx4$\% to $\approx8$\%) with $\Sigma_{SFR}$ (in the range between 0.0005 and 0.003$~M_{\odot}~\mbox{yr}^{-1}~\mbox{kpc}^{-2}$),
among eight regions within M31 from the PHAT survey (Dalcanton et al. 2012; Williams et al. 2014).
%Their results for $\Gamma_S(10,100)$ in M31 support a modest increase, from $\approx 4$\% to $\approx8$\%, over a range of $0.0005$ to 0.003 in $\Sigma_{SFR}$, but with a significant outlying point.
It seems quite plausible that there are variations in $\Gamma_S(10,100)$ (as well as in $\Gamma_S(0,10)$ and $\Gamma_S(100,400)$) on the smaller physical scales studied by Johnson et al. (2016), because similar variations have been found for the cluster age and mass distributions in kpc-scale regions of other galaxies (e.g., Chandar et al. 2014; Adamo et al. 2015).
%Our study only includes two galaxies (the LMC and SMC) within this range of $\Sigma_{SFR}$, and does not include an analysis of subregions within a single galaxy.  
We note that while the results for $\Gamma$ presented in Johnson et al. (2016) for M31 are based on homogeneous procedures, their compilation of results for other galaxies suffers from the same selection bias and heterogeneous procedures shown here in Figure~1.

The observed shapes of the CMFs provide additional constraints on disruption processes.
The older mass functions plotted in Figures~4 and 7 do not show any significant flattening at the low mass end, as would be expected if lower mass clusters were disrupted earlier than higher mass clusters (e.g., Fall et al. 2009).
Therefore, the disruption rates of clusters must be approximately independent of their masses, at least over the mass-age ranges studied here.
%In Section~5, we showed that the typical $\Gamma_S$ values are similar to the predictions from a $\gamma=-0.7$ disruption model, i.e. one where approximately 80\% of the clusters dissolve each decade in age, regardless of their initial mass.
Fall \& Chandar (2012) review the primary disruption mechanisms for star clusters and explain why their rates are independent of mass or nearly so.

One of the mechanisms that can disrupt star clusters is tidal interactions with passing giant molecular clouds (GMCs: Spitzer 1958; Binney \& Tremaine 2008).
In this context, it may seem puzzling that the values of $\Gamma_S(10,100)$ and $\Gamma_S(100,400)$ show little or no dependence on the surface density of molecular gas $\Sigma_{H_2}$.
There are two possible reasons for this.
First, in our sample of eight galaxies, $\Sigma_{H_2}$ varies by only a factor of $\sim10$, thus making any trend difficult to detect.
Second, the characteristic time-scale $\tau_d$ for disruption by this mechanism may be too long to show up in the age intervals examined here, $\tau < 4\times10^8$~yr.
For clusters in the solar neighborhood, the disruption timescale is $\tau_d\sim3\times10^8$~yr, with uncertainties of at least a factor of 2 (Binney \& Tremaine 2008).
For clusters in other galaxies, the uncertainties are much greater, because $\tau_d$ depends in general on the characteristic internal densities of the clusters $\rho_h$ and the mass spectrum, mass-radius relation, number density, and velocity dispersion of the GMCs.
Unfortunately, we do not have reliable estimates of most of these quantities outside the Milky Way.
In a particular limit, the catastrophic regime, $\tau_d$ depends only on $\rho_h$ and $\Sigma_{H_2}$; but in order to determine whether this case is applicable, rather than the opposite limit, the diffusive regime, one must know the values of all the other quantities listed above (see Binney \& Tremaine 2008 for a clear and thorough analysis).
Thus, we simply do not have enough information about the properties of clusters and GMCs in other galaxies to estimate $\tau_d$ reliably.

Our new results for $\Gamma$ and CMF$/$SFR support the quasi-universal model (e.g., Whitmore et al. 2007; Fall \& Chandar 2012; Chandar et al. 2014, CFW15), which postulates that clusters in different galaxies form and disrupt in similar ways, with relatively minor variations within and among galaxies. 
The ''initial'' mass functions of young ($\tau \lea 10$~Myr) star clusters have a nearly universal power-law shape, $dN/dM \propto M^{\beta}$, with $\beta\approx-2$ (see Figure~1 of Fall \& Chandar 2012 and Figure~4 here).
Here, we have shown that the normalization of this power law is set by the overall star formation rate of the host galaxy, with a similar $\approx24$\% of stars forming in compact clusters in different galaxies.
Because the shapes of the mass functions do not change much with age, the disruption of the clusters must be roughly independent of their initial masses 
(e.g., Fall et al. 2005; Chandar et al. 2010; Fall \& Chandar 2012).
The rapid decline in the age distributions, $dN/d\tau \propto \tau^{\gamma}$ with $\gamma\approx-0.7$ for $\tau \gea 10$~Myr, the signature of this disruption, is also reflected in the differences we find between $\Gamma_F$ and $\Gamma_S(10,100)$ and $\Gamma_S(100,400)$.
The joint distribution of masses and ages for the quasi-universal model for cluster formation and disruption can then be written in compact form as
$g(M,\tau) = c \times \mbox{SFR} \times M^{\beta}~\tau^{\gamma}$, with $c\approx0.24$, $\beta\approx-2$ and $\gamma\approx-0.7$. 
This appears to apply, at least approximately, to the cluster populations in
star-forming galaxies over a wide range of conditions.
We do expect small, second-order regional variations in the properties of the clusters within galaxies, but larger, uniform cluster samples will be needed in order to detect these.

\section{SUMMARY}

In this work, we compiled catalogs of star clusters in eight different
galaxies (the LMC, SMC, NGC~4214, NGC~4449, M83, M51, the Antennae,
and NGC~3256), which includes irregulars, dwarf starbursts, spirals and mergers 
that span a fairly broad range in star-forming properties.
%$\Sigma_{SFR}$, SFR, and $\Sigma_{H_2}$.  
For these eight galaxies, we 
measured consistently the fraction of stars that {\em form} in compact clusters, $\Gamma_F$, and the fraction of stars in clusters that {\em survive} to older ages, $\Gamma_S(\tau_1,\tau_2)$.  
We also compared the mass functions of the clusters, before and after dividing by the star formation rates (the CMF$/$SFR method), as a check on our new results for $\Gamma_F$ and $\Gamma_S$.  
The main conclusions of this paper are the following:

\begin{itemize}
\item The typical fraction of stars that form in compact clusters, bound and unbound, is $\Gamma_F\approx\Gamma_S(0,10)\approx24\pm9$\%, for galaxies in our sample.  
This is confirmed by a more detailed study of clusters in the LMC.
%Including the impact of various assumptions and systematic effects, 
We estimate that this result is uncertain by no more than a factor of $\approx1.7$.
%This value can vary by a factor of {\bf $\approx 1.5$}, depending on the specific assumptions that are made in the calculation.

\item The fraction of stars in surviving clusters declines with age: $\Gamma_S(10,100)=4.6\pm2.5$\% and $\Gamma_S(100,400)=2.4\pm1.1$\%.
These values support a picture in which $\approx70-80$\% of the clusters disrupt in each decade in age, with similar results among different galaxies.

\item Our new results for $\Gamma_F$ and $\Gamma_S$ are similar to those from the CMF$/$SFR method.
Neither method shows any significant dependence on $\Sigma_{SFR}$, SFR, or $\Sigma_{H_2}$ of the host galaxy.
This appears to contradict previously published results, which have claimed a strong increase in $\Gamma_F$ with $\Sigma_{SFR}$ and $\Sigma_{H_2}$.
However, the previous results
%that claimed a strong increase in the fraction of stars born in clusters with 
were biased in such a way that
older (younger) clusters were used to calculate $\Gamma_S$ ($\Gamma_F$) in galaxies with lower (higher) $\Sigma_{SFR}$, which then resulted in the apparent trends.

\item An important conclusion from this paper is that future studies that compare cluster populations among different galaxies should adopt a consistent set of assumptions and procedures.

\end{itemize}

{\it Facilities:} \facility{HST}.
\acknowledgments We thank the anonymous referee for suggestions that improved this paper.  R. C. acknowledges support from NSF grant 1517819.  S. M. F. appreciates the hospitality of the Kavli Institute for Theoretical Physics and the Aspen Center for Physics, which are supported in part by NSF grants PHY11-25915 and PHYS-1066293.

\begin{figure}
\plotone{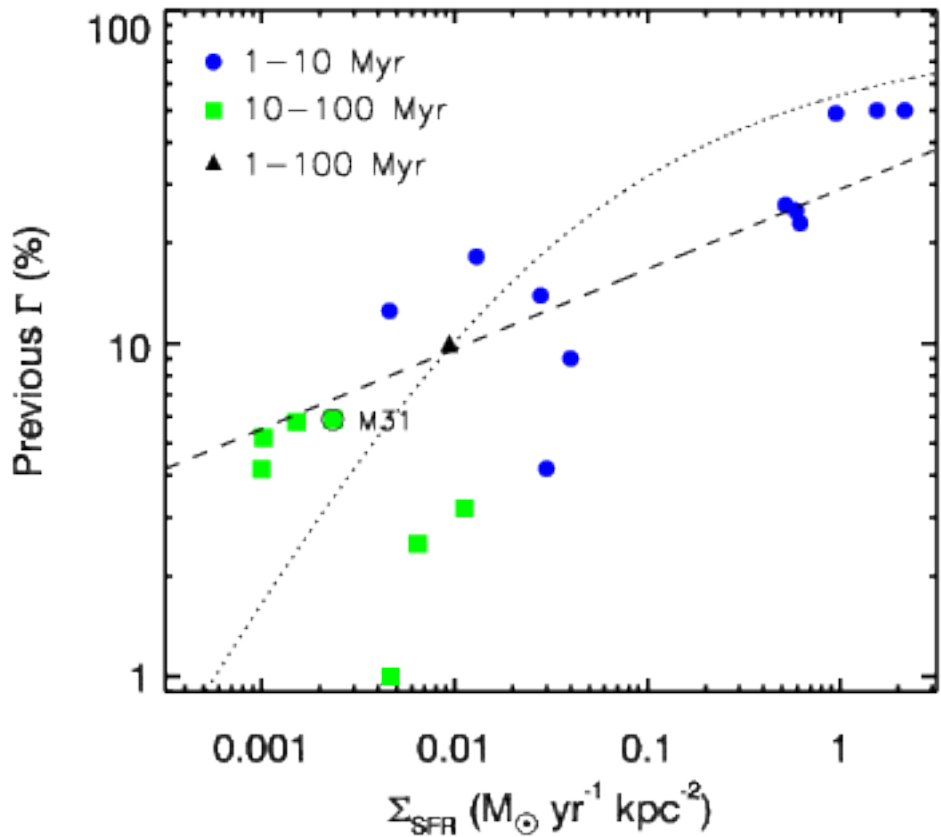}
  \caption{Previously derived fractions of stars in clusters in different age intervals (compiled by Adamo et al. 2015) plotted against the mean surface density of SFR in the host galaxy.
The average result for M31 from Johnson et al. (2016) is also shown.
This plot shows that different age ranges have been used for different galaxies, highlighted by the different color symbols.
There has been a tendency in the literature to use the $10-100$~Myr age interval for galaxies with lower $\Sigma_{SFR}$ and the $1-10$~Myr age interval for galaxies with higher $\Sigma_{SFR}$.
% and to directly compare the results for the two different quantities.
This has led to a steep apparent trend.
% that has been highly cited in the literature.
The dashed line represents the empirical relationship determined by Goddard et al. (2010) from 6 galaxies which had a mix of intervals used to determine $\Gamma$, and the dotted curve shows the predicted relationship from Kruijssen (2012).}
\label{fig:gamma}
\end{figure}

\begin{figure}
% \plotone{figcompsfrs}
 \plotone{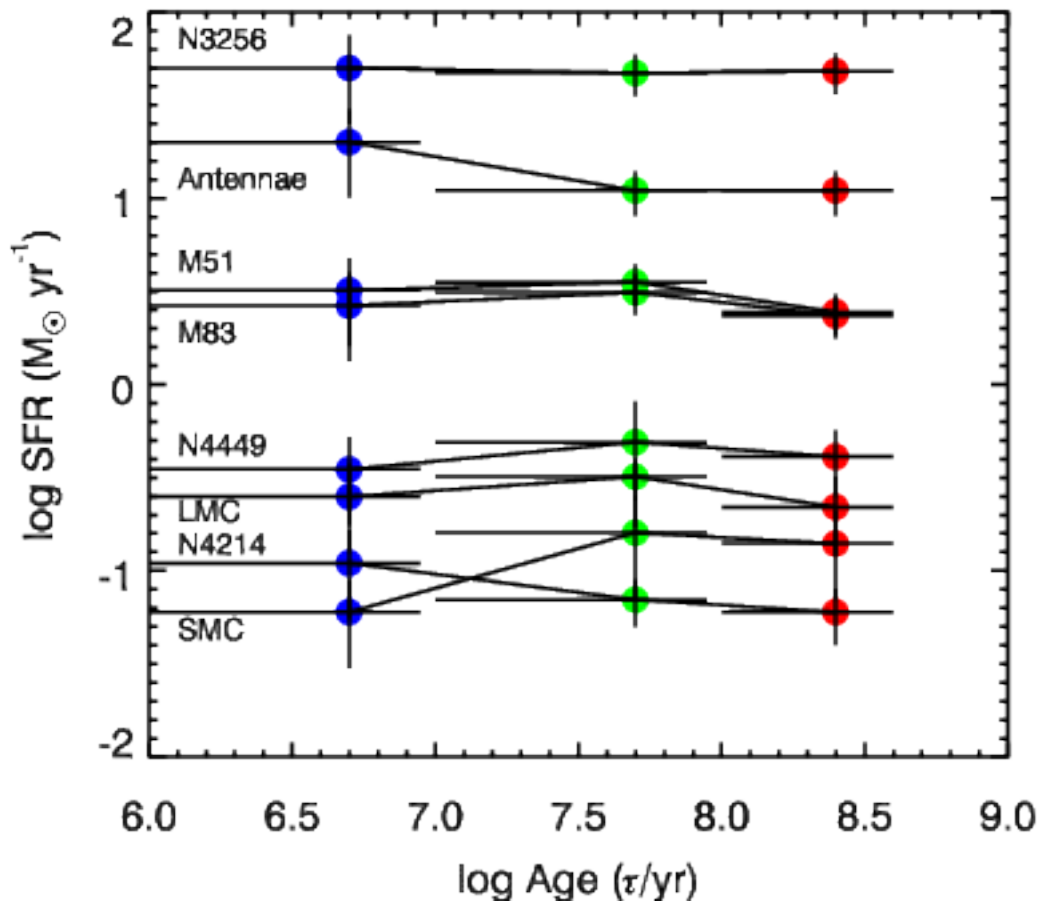}
  \caption{The average star formation rates compiled in Table~2 are plotted for the three age intervals of interest: (1) $< 10$~Myr, (2) $10-100$~Myr, and (3) $100-400$~Myr.  For the $<10$~Myr interval, we use our newly determined SFRs from extinction-corrected H$\alpha$. 
For the LMC, SMC, NGC~4214, and NGC~4449 (the 4 closest galaxies),
we determine the average SFR from published CMD analysis within the $10-100$~Myr and $100-400$~Myr intervals.
For the four more distant galaxies, we adopt our extinction-corrected FUV values for the $10-100$~Myr interval and the IR-based values for the $100-400$~Myr interval. 
This figure demonstrates that the SFRs have not varied by more than a factor of $\sim2$ for galaxies in our sample, and more importantly, that there are no systematic trends in the average SFR with age.
}
  \label{fig:sfrs}
\end{figure}

\begin{figure}
\epsscale{0.7}
% \plotone{../Figures/dndt.eps}
 \plotone{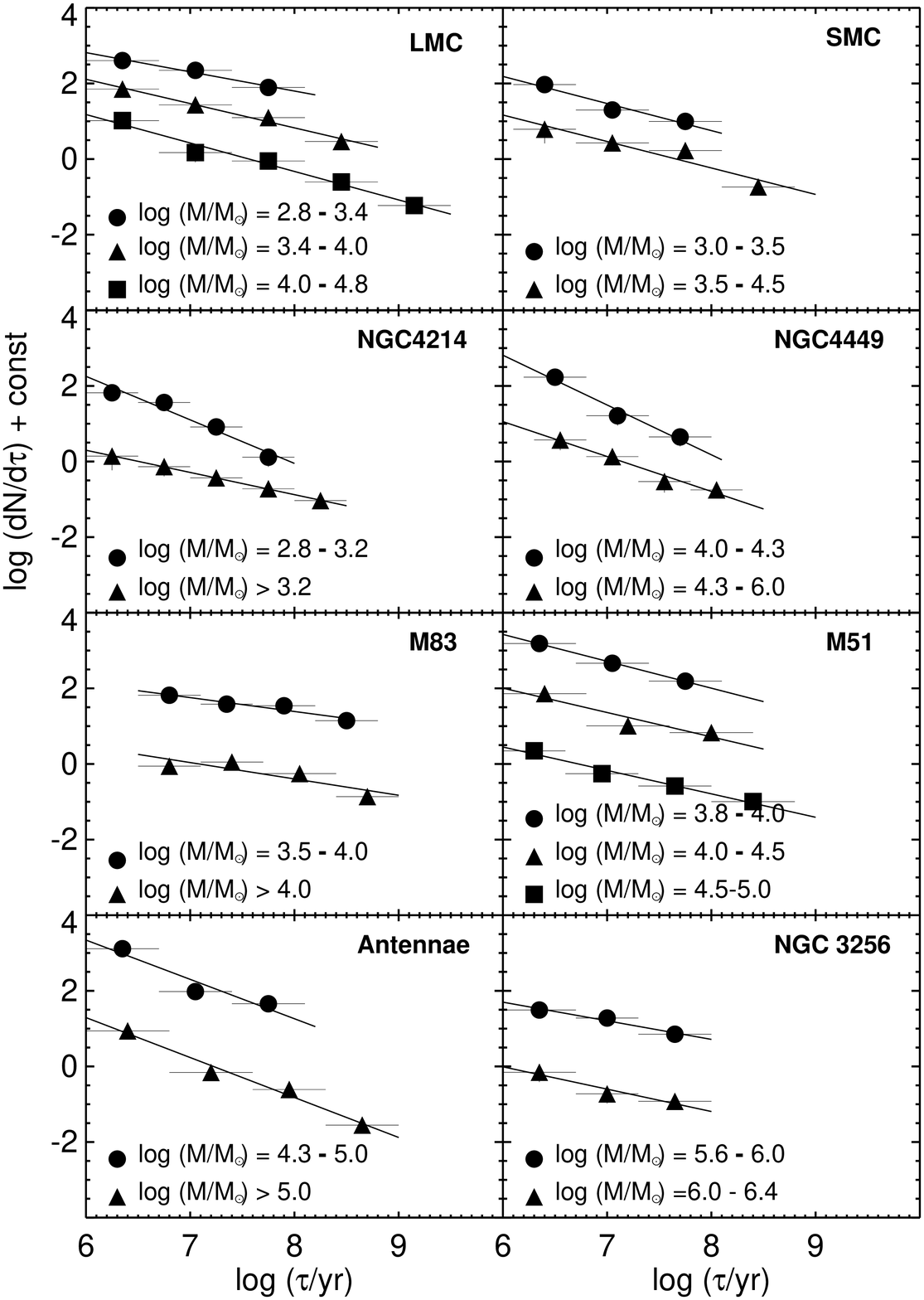}
  \caption{Age distributions of star clusters in our sample galaxies in the indicated mass intervals.
The lines show power-laws,  $dN/d \tau \propto \tau^{\gamma}$, with the best-fitting exponents $\gamma$ listed in Table~3. 
All of these results have been published previously, except for NGC~4214 and M83.  For M83, our results here supercede the earlier one by Chandar et al. (2014), because we now use a cluster catalog based on seven pointings with the {\em Hubble Space Telescope}, rather than just two.  
}
  \label{fig:dndt}
\end{figure}

\begin{figure}
\plotone{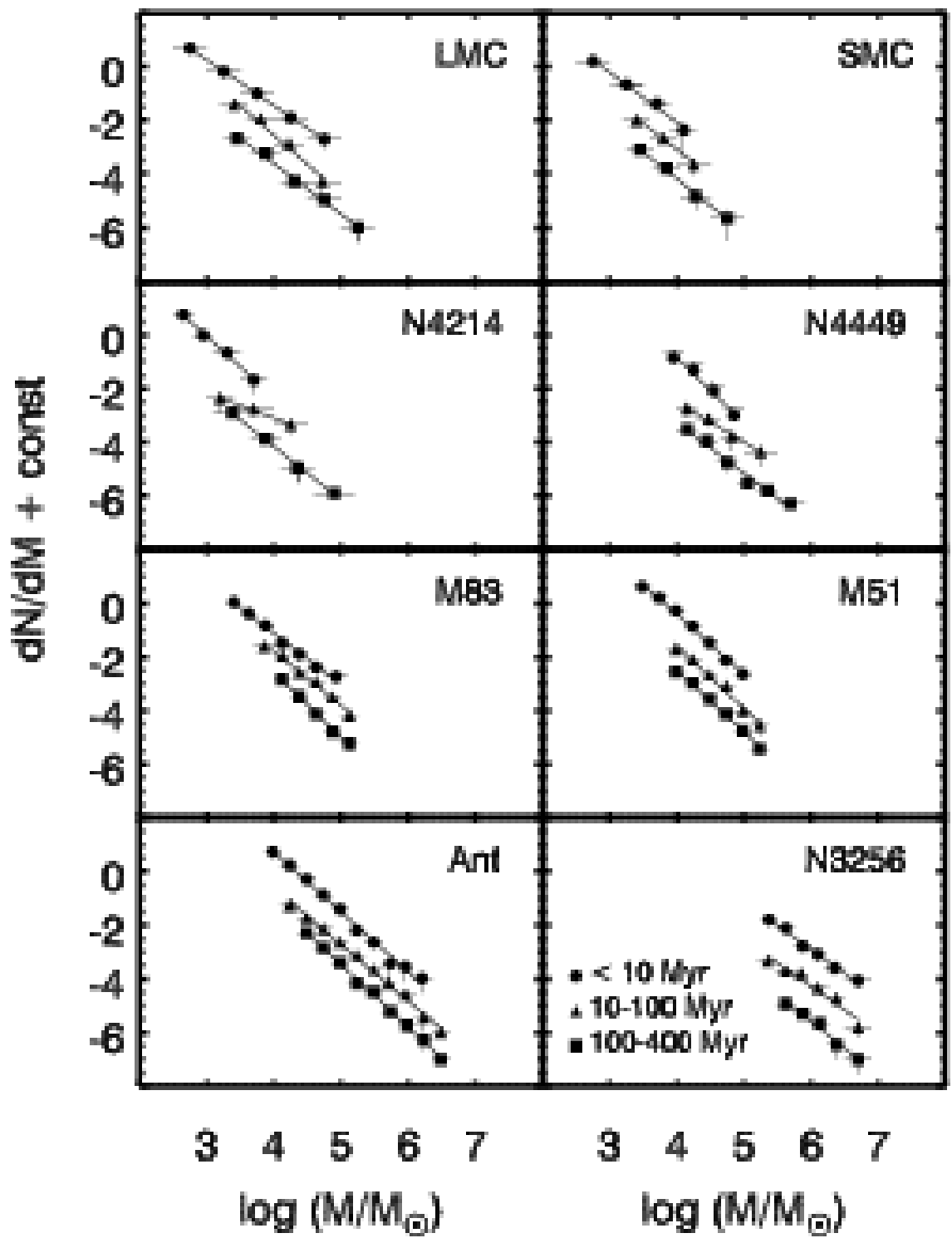}
\caption{Mass functions of star clusters in our sample galaxies in three age intervals: $\tau < 10$~Myr (circles), $\tau=10-100$~Myr (triangles), and $\tau=100-400$~Myr (squares).
The lines show power-laws, $dN/dM \propto M^{\beta}$, with the best fitting exponents, most of which are close to $\beta=-2.0$.
}
\label{fig:dndm}
\end{figure}

\begin{figure}
% \plotone{../Figures/f2new}
 \plotone{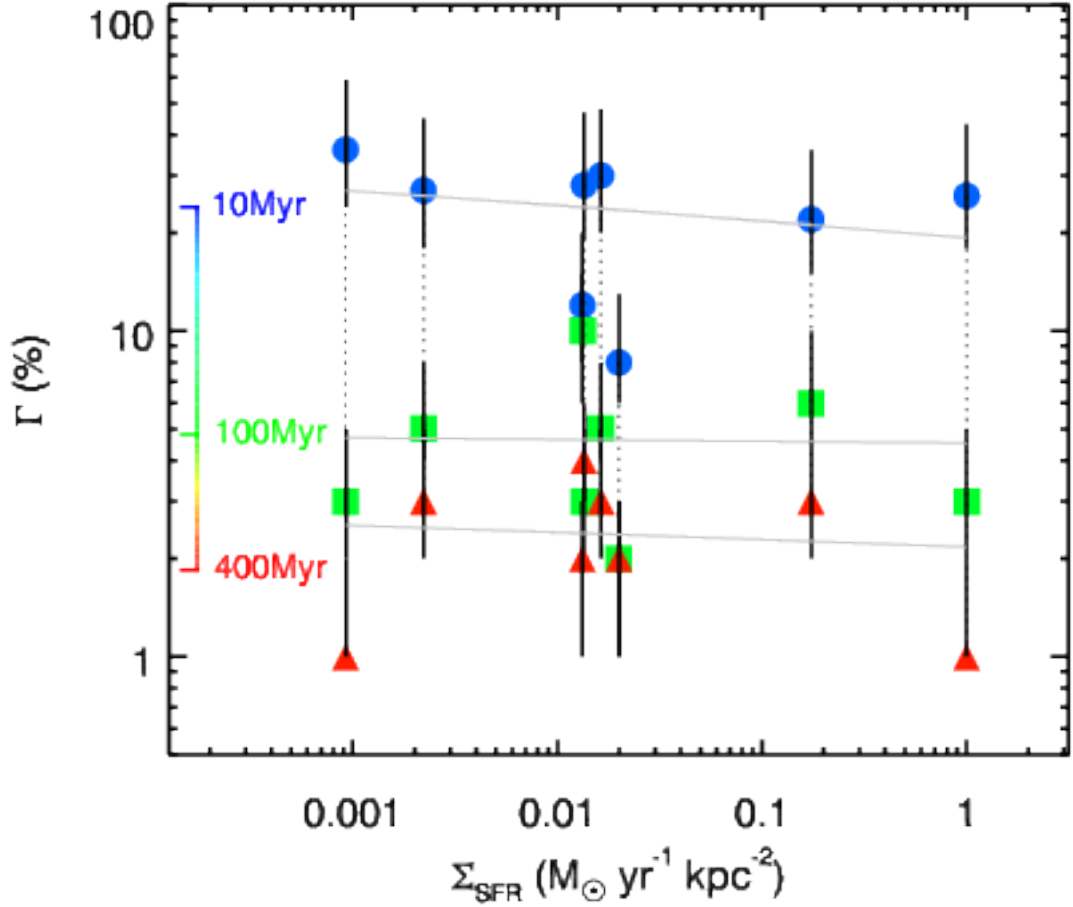}
  \caption{Some of the main results of this work.  Our new determinations of $\Gamma_F$ (blue circles) for eight galaxies show that $\approx24$\% of stars form in clusters, and that this fraction does not vary systematically with $\Sigma_{SFR}$ of the host galaxy (the best fit is shown as the gray line).
Our calculations for $\Gamma_S(10,100)$ (green squares) and $\Gamma_S(100,400)$ (red triangles) are also plotted for the same galaxies, and show that there is a systematic decrease in the fraction of stars in surviving clusters with increasing age, a reflection of cluster disruption.
The color bar on the left shows that the predicted decrease from a $\gamma=-0.7$ disruption model, starting at $\Gamma=24$\% (the mean $\Gamma_F$ value determined for this sample), provides a good match to our observational results.
}
  \label{fig:moneyplot}
\end{figure}

\begin{figure}
% \plotone{../Figures/f3}
 \plotone{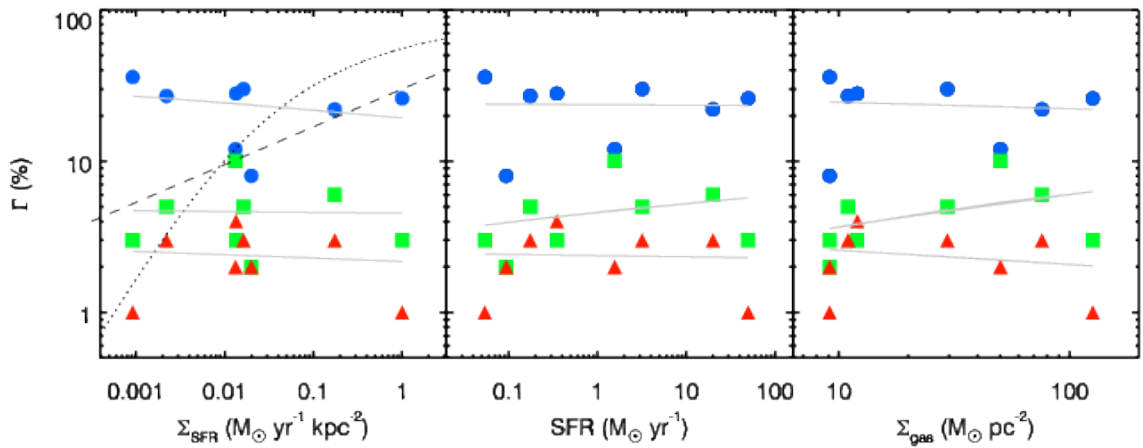}
  \caption{Our new determinations of $\Gamma_F$ (blue circles), $\Gamma_S(10,100)$ (green squares), and $\Gamma_S(100,400)$ (red triangles) are plotted against $\Sigma_{SFR}$, SFR, and $\Sigma_{H_2}$ for our sample galaxies.
No statistically significant correlations are found between any pair of plotted quantities. 
In the left panel, the dashed line shows the empirical relation between $\Gamma$ and $\Sigma_{SFR}$ presented by Goddard et al. (2010) 
and the curved, dotted line shows the even stronger relation predicted by the Kruijssen (2012) model.
Neither one of these provide an acceptable fit to our new results.
}
  \label{fig:newgamma}
\end{figure}

\begin{figure}
\epsscale{0.75}
\plotone{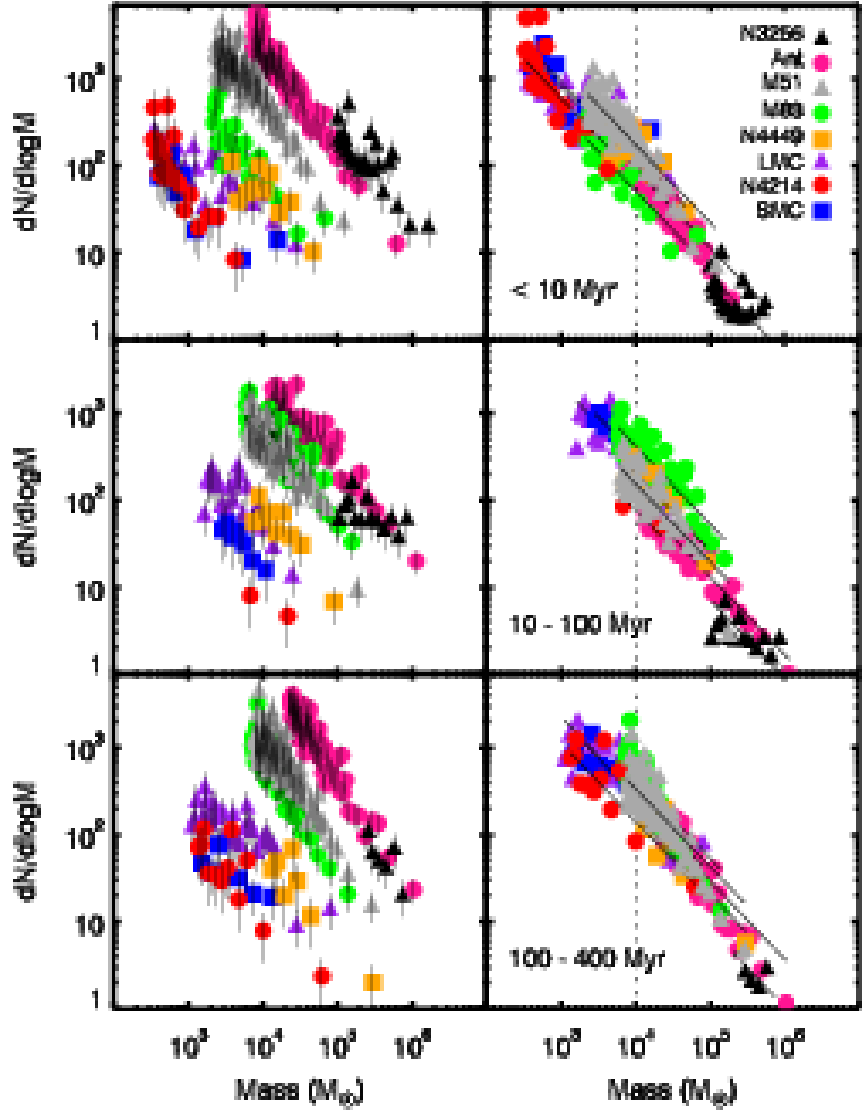}
  \caption{Observed mass functions of clusters in our sample galaxies are presented in the left panels in the given interval of age.
The right panels show that when the CMFs are divided by the SFRs of their host galaxies, the resulting CMF$/$SFR distributions lie very close to each other. 
The lines show the best fit from equation~(4) in Chandar et al. (2015), which is used to determine the amplitude $A$ of each CMF at $M=10^4~M_{\odot}$.}
  \label{fig:mf}
\end{figure}

\begin{figure}
 \plotone{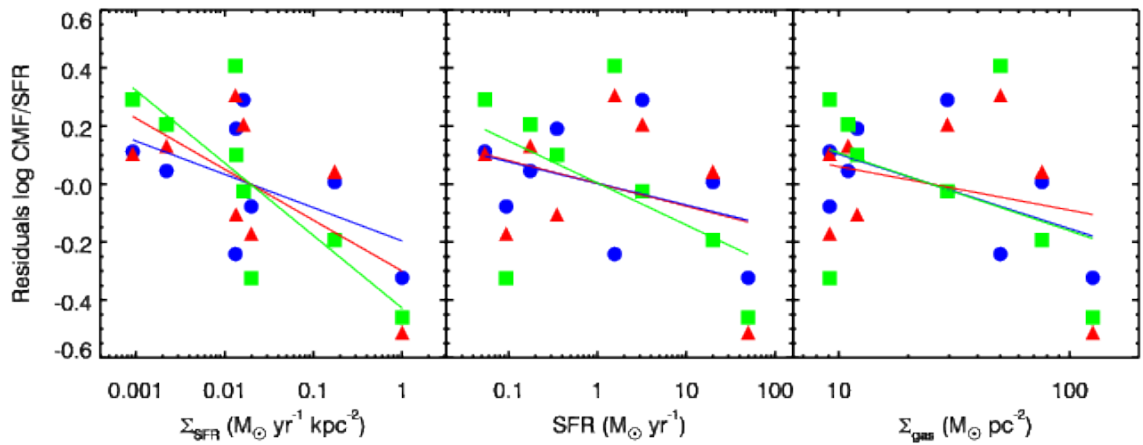}
 \caption{Logarithmic residuals in the amplitudes of the CMF$/$SFR distributions (denoted by $A$ in the text)
are plotted against $\Sigma_{SFR}$, SFR, and $\Sigma_{H_2}$ for clusters in the age intervals: $< 10$~Myr (blue circles), $10-100$~Myr (green squares) and $100-400$~Myr (red triangles).
No statistically significant correlations are found between any pair of plotted parameters.
There is a very weak, $\sim2\sigma$ anti-correlation between the residuals for $10-100$~Myr clusters and $\Sigma_{SFR}$, but this is consistent with observational uncertainties alone.
% are likely due to uncertainties in the results rather than any physical trends.
 }
  \label{fig:cmfsfr}
\end{figure}

%\clearpage

\clearpage

\begin{figure}
\plotone{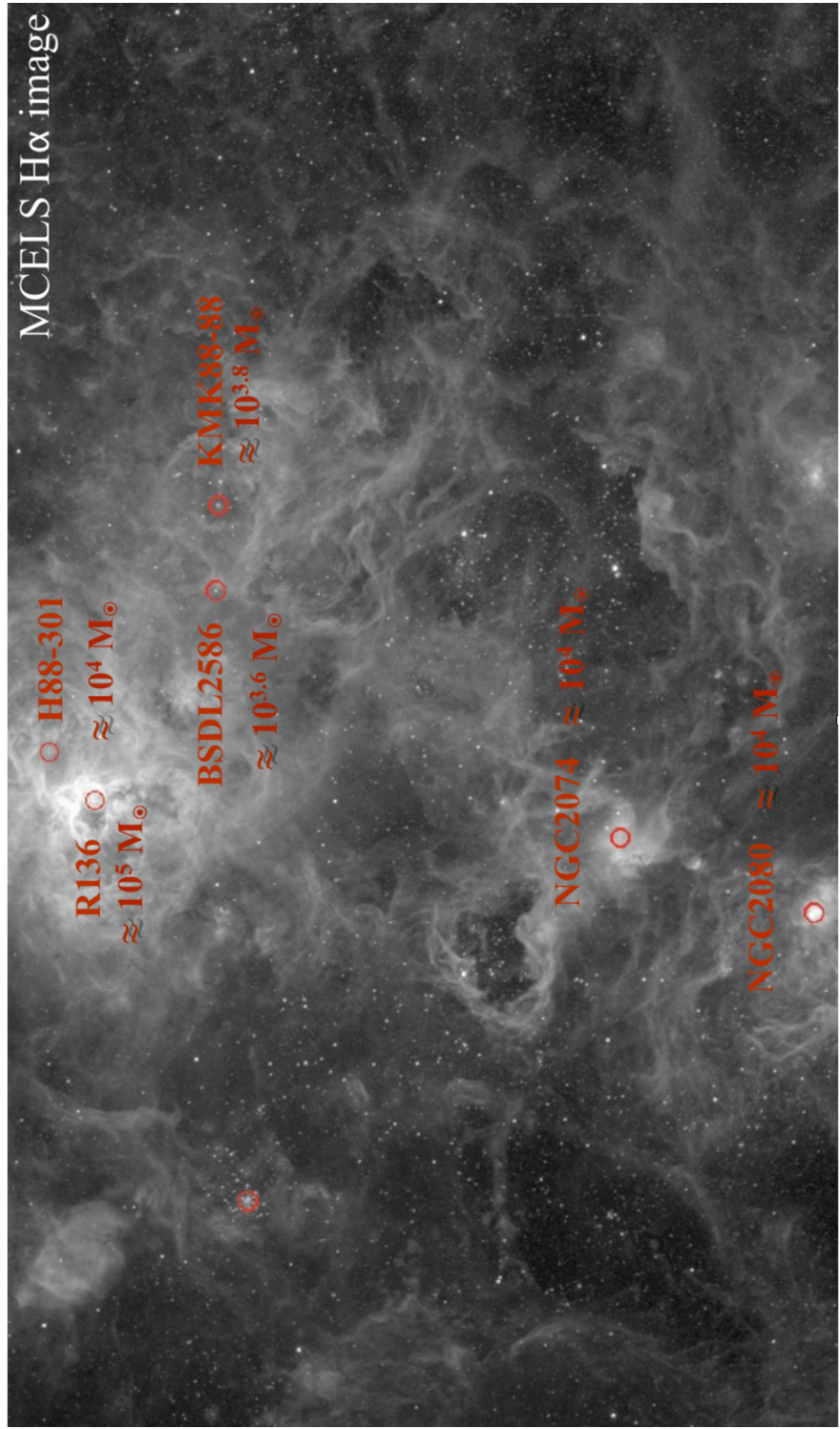}
  \caption{An H$\alpha$ image of a $1 \times 2$~kpc region of the LMC which has formed a number of massive, young clusters ($\tau < 10$~Myr).  A significant portion of the recent star formation appears to have occurred in this type of compact clusters.}
  \label{fig:mf}
\end{figure}

%\clearpage
% **************  TABLE 1 :  GALAXY PROPERTIES ****************

% \pagestyle{empty}
\begin{rotatetable}
  \begin{deluxetable}{ccccccccccc}
    \tablecolumns{3} \tablecaption{Properties of Sample Galaxies\label{galaxies}} \tablewidth{0pt} \tablehead{
%      \colhead{}  & \colhead{}  & \colhead{}  & \colhead{H$\alpha$ + 25$\mu$m} & \colhead{SFR density} & \colhead{Gas Density} & \colhead{SFR} \\
      \colhead{Galaxy}   & \colhead{Distance}  & \colhead{SFR} & \colhead{\%} & \colhead{Area} & \colhead{$\Sigma_{SFR}$ (this work)} &
\colhead{$\Sigma_{SFR}$ (literature)} &
 \colhead{$\Sigma_{H_2}$}   \\
      \colhead{Name}  & \colhead{(Mpc)}  & \colhead{($M_{\odot}$yr$^{-1}$)} & \colhead{Coverage} & \colhead{(kpc$^{2}$)} & \colhead{($M_{\odot}$~yr$^{-1}$~kpc$^{-2}$)} & 
\colhead{($M_{\odot}$~yr$^{-1}$~kpc$^{-2}$)} & 
\colhead{($M_{\odot}$~pc$^{-2}$)}  \\ \hline } 
\startdata
    LMC   &  0.05    & 0.25 & 70 & 79 & $2.21\times10^{-3}$ & $1.52\times10^{-3}$ &   11 \\
    SMC  & 0.06  &  0.06 & 90 & 43 & $9.22\times10^{-4}$  & 0.001  &   9 \\
    NGC 4214  &  3.1 &  0.11 & 100 & 5  & $0.022$ & $3.80\times10^{-3}$  &   9.2 \\
    NGC 4449  & 3.8  &  0.35 &  85 & 12 & $0.013$  & 0.04  &    12 \\
    M83  & 4.5 &  2.65 & 60 & 75 & 0.013 & $0.013$  &  50 \\
    M51 & 8.2 & 3.20  & 90 & 196 & $0.016$  &  $8.21\times10^{-3}$   &   30 (53) \\
    Antennae & 22  &  20 & 100 & 113  & 0.174  & ...   & 76 (125) \\
    NGC~3256 & 36 & 50 & 100 & 50 & 1.0  & 0.62 &   125 \\
% previous coverage, in order: 79, 43, 21,25,166,578,150,74
 \enddata
 %\tablenotetext{1}{Power-law index for the mass function of star clusters, determined from the least-squares fits to log$(dN/d\mbox{log}M) = (\beta-1)\log{M} + {\rm const}$}
%\tablenotetext{2}{Star formation rates determined in this work.  See Section~3.1 for a description} 
\tablecomments{The literature $\Sigma_{SFR}$ values are taken from the compilation in Adamo et al. (2015). The $\Sigma_{H_2}$ values are taken from the compilation in Kruijssen \& Bastian (2016), with updated values for M51 and the Antennae, described in Section~3.3, given in parentheses.}
%References: 1. James et al. 2008; 2. Harris \& Zaritsky 2009; 3. Bolatto et al. 2011; 4. Harris \& Zaritsky 2004); 5. Lee et al. 2009; 6. Zhang et al. 2001 }
    % \tablenotetext{1}{Least-squares fits to log$(dN/dM) = \beta\log{M} + {\rm const}$}
  \end{deluxetable}
\end{rotatetable}
% Leroy et al. 2013 give 9.2 for density of H_1 + H_2 gas in N4214
% Leroy et al. 2013 give 53 for density in M51 (N5194), but KB16 give 30 !
% Zhang, Fall, Whitmore give: 1.5e10 Msun of H2 gas -> 1.5e10/(120.d6)

% ****** TABLE 2: SFR DETERMINATIONS *****

% \pagestyle{empty}
\begin{rotatetable}
  \begin{deluxetable}{ccccccccccc}
    \tablecolumns{3} \tablecaption{Comparison of Different Star Formation Rate Determinations\label{galaxies}} \tablewidth{0pt} \tablehead{
%      \colhead{}  & \colhead{}  & \colhead{}  & \colhead{H$\alpha$ + 25$\mu$m} & \colhead{SFR density} & \colhead{Gas Density} & \colhead{SFR} \\
      \colhead{Galaxy}   & \colhead{H$\alpha$ SFR}  & \colhead{FUV SFR} & \colhead{24$\mu$m SFR} & \colhead{CMD SFR} & \colhead{CMD SFR}  \\
      \colhead{Name}   & \colhead{($M_{\odot}$yr$^{-1}$)} & \colhead{($M_{\odot}$yr$^{-1}$)} & \colhead{($M_{\odot}$yr$^{-1}$)} & \colhead{($M_{\odot}$yr$^{-1}$)}  & \colhead{$M_{\odot}$yr$^{-1}$)} \\ 
\colhead{} & \colhead{($\sim1-10$~Myr)} & \colhead{($\sim1-100$~Myr)} & \colhead{($\sim1-400$~Myr)} & \colhead{($10-100$~Myr)} & \colhead{($100-400$~Myr)} \\ \hline } 
\startdata
    LMC   & $0.25\pm0.12$ & ... & ... & $0.32\pm0.08$ (1) & $0.22\pm0.10$ (1) \\ 
    SMC  & $0.06\pm0.03$ & ... & ... & $0.16\pm0.09$ (2)  & $0.14\pm0.07$ (2) \\ 
    NGC 4214  & $0.11\pm0.05$ & $0.16\pm0.04$ & $0.06\pm0.02$  & $0.07\pm0.02$ (3) & $0.06\pm0.02$ (3) \\ 
    NGC 4449  & $0.35\pm0.18$ & $0.42\pm0.10$ & $0.20\pm0.05$ & $0.49\pm0.32$ (3) & $0.41\pm0.16$ (3) \\ 
    M83  & $2.65\pm1.3$ & $3.12\pm0.78$ & $2.35\pm0.59$ & $>1.20$ & ... \\ 
    M51 & $3.20\pm1.6$ & $3.54\pm0.89$ & $2.45\pm0.61$ & ... & ... \\ 
    Antennae & $20\pm10$ & $11\pm3$ & $11\pm3$ & ... & ... \\ 
    NGC~3256 & $50\pm17$ & $47\pm12$ & $48\pm12$ & ... & ... \\  \hline
 \enddata
\tablecomments{CMD-based SFRs are difficult to determine in galaxies more distant than $\sim8$~Mpc. FUV measurements from GALEX do not exist for the LMC and SMC. CMD-based SFRs are determined over the last 100~Myr, unless otherwise noted.  {\bf References:} 1. Harris \& Zaritsky 2004; 2. Harris \& Zaritsky 2009; 3. McQuinn et al. 2010. }
  \end{deluxetable}
\end{rotatetable}

% ***********TABLE 3: Exponenets of Age Distributions ****************

%\clearpage
% \pagestyle{empty}
\begin{rotatetable}
  \begin{deluxetable}{ccccccccccc}
    \tablecolumns{3} \tablecaption{Exponents of Age Distributions} \tablewidth{0pt} \tablehead{
\colhead{Galaxy} & \colhead{Mass Interval} & \colhead{$\gamma$\tablenotemark{a}} \\
\colhead{Name} & \colhead{log~($M/M_{\odot}$)} & \colhead{} \\ \hline } 
\startdata
    LMC  & $4.0-4.8$  & $-0.75\pm0.07$  \\
    LMC  & $3.4-4.0$  & $-0.68\pm0.07$  \\
    LMC  & $2.8-3.4$  & $-0.54\pm0.10$  \\
    SMC  & $3.5-4.5$  & $-0.73\pm0.17$  \\
    SMC  & $3.0-3.5$  & $-0.70\pm0.16$  \\
    NGC 4214  & $>3.2$   & $-1.15\pm0.18$  \\
    NGC 4214  & $2.8-3.2$   & $-0.59\pm0.02$  \\
    NGC 4449  & $4.3-6.0$   & $-1.30\pm0.22$  \\
    NGC 4449  & $4.0-4.3$   & $-0.92\pm0.12$  \\
    M83  & $>4.0$   & $-0.37\pm0.08$  \\
    M83  & $3.5-4.0$   & $-0.43\pm0.17$  \\
    M51  & $4.5-6.0$  & $-0.62\pm0.07$  \\
    M51  & $4.0-4.5$  & $-0.64\pm0.20$  \\
    M51  & $3.8-4.0$  & $-0.71\pm0.03$  \\
    Antennae & $>5.0$   & $-1.05\pm0.11$  \\
    Antennae & $4.3-5.0$   & $-1.04\pm0.30$  \\
    NGC~3256  & $6.0-6.4$  & $-0.59\pm0.17$ \\ 
    NGC~3256  & $5.6-6.0$  & $-0.49\pm0.10$ \\  \hline
 \enddata
\tablenotetext{a}{ Least-squares fits to log~($dN/d\tau) = \gamma~\mbox{log}\tau + \mbox{const.}$}
  \end{deluxetable}
\end{rotatetable}

% ********** TABLE 4: VALUES OF Mlim *************************

%\clearpage
% \pagestyle{empty}
\begin{rotatetable}
  \begin{deluxetable}{ccccccccccc}
    \tablecolumns{3} \tablecaption{Lower Limits of Cluster Masses $M_{lim}$ Used to Calculate $\Gamma$} \tablewidth{0pt} \tablehead{
\colhead{Galaxy} & \colhead{log~$M_{lim}$} & \colhead{log~$M_{lim}$} & \colhead{log~$M_{lim}$} \\
\colhead{Name} & \colhead{$<10$~Myr ($M_{\odot}$)} & \colhead{$10-100$~Myr ($M_{\odot}$)} & \colhead{$100-400$~Myr ($M_{\odot}$)} \\ } 
\startdata
    LMC  & 2.5  & 3.2  & 3.25  \\
    SMC  & 2.5   & 3.2  & 3.25  \\
    NGC 4214  & 2.5   & 3.0  & 3.0  \\
    NGC 4449  & 3.4   & 4.0  & 4.0   \\
    M83  & 3.3    & 3.7  & 4.0  \\
    M51  & 3.5  & 3.9  & 4.0   \\
    Antennae & 4.0   & 4.25  & 4.5  \\
    NGC~3256  & 5.2  & 5.2  & 5.5  \\ \hline
 \enddata
%\tablecomments{
%The uncertainties reflect uncertainties in the total cluster mass and in the SFR (as discussed in Section~3), and are assumed to be 10\% for the very youngest clusters used to calculate $\Gamma_F$ and 4\% for the older clusters used to calculate $\Gamma_S$.  We assume that the SFR is uncertain by 50\% for all calculations. }
  \end{deluxetable}
\end{rotatetable}

% ********** TABLE 4: OUR GAMMA VALUES *************************

%\clearpage
% \pagestyle{empty}
\begin{rotatetable}
  \begin{deluxetable}{ccccccccccc}
    \tablecolumns{3} \tablecaption{New Determinations of $\Gamma_F$ and $\Gamma_S$\label{gamma}} \tablewidth{0pt} \tablehead{
\colhead{Galaxy} & \colhead{$\Gamma_F$} & \colhead{$\Gamma_S$ (10,100)} & \colhead{$\Gamma_S$(100,400)} \\
\colhead{Name} & \colhead{(\%)} & \colhead{(\%)} & \colhead{(\%)} \\ \hline } 
\startdata
    LMC  & $27^{+18}_{-9}$ & $5^{+3}_{-1}$ & $3^{+2}_{-1}$ \\
    SMC  & $36^{+23}_{-12}$ & $3^{+2}_{-1}$ & $1^{+1}$   \\
    NGC 4214  & $8^{+5}_{-2}$ & $2\pm1$ & $2\pm1$ \\
    NGC 4449  & $28^{+19}_{-9}$ & $3^{+2}_{-1}$ & $4^{+2}_{-1}$ \\
    M83  &  $12^{+8}_{-4}$ & $10^{+5}_{-4}$ & $2\pm1$  \\
    M51  & $30^{+18}_{-10}$ & $5^{+3}_{-1}$ & $3^{+2}_{-1}$ \\
    Antennae & $22^{+14}_{-7}$ & $6^{+4}_{-1}$ & $3^{+2}_{-1}$   \\
    NGC~3256  & $26^{+17}_{-8}\tablenotemark{a}$ & $3^{+2}_{-1}$ & $1^{+1}$ \\ \hline
    Mean, $\sigma$ & $24\pm9$ & $4.6\pm2.5$ & $2.4\pm1.1$ \\ 
    Median &  27   &  5  & 3  \\ \hline
 \enddata
\tablenotetext{a}{The value of $\Gamma_F$ determined here for NGC~3256 is 22\% lower than that determined in Mulia et al. (2016), because we have made a more conservative correction, $10/9$ rather than $10/7$, for highly obscured, missing clusters.
See Section~5.1 for details.}
%\tablecomments{
%The uncertainties reflect uncertainties in the total cluster mass and in the SFR (as discussed in Section~3), and are assumed to be 10\% for the very youngest clusters used to calculate $\Gamma_F$ and 4\% for the older clusters used to calculate $\Gamma_S$.  We assume that the SFR is uncertain by 50\% for all calculations. }
  \end{deluxetable}
\end{rotatetable}

% ***********TABLE 6:  PREVIOUS DETERMINATIONS OF GAMMA ****************

\begin{rotatetable}
\begin{deluxetable}{ccccccccccc}
%  \tablecolumns{2} 
\tabletypesize{\footnotesize}
\tablecaption{Previous Determinations and Assumptions for $\Gamma_F$ and $\Gamma_S$\label{tab:fits}} 
\tablewidth{0pt} 
\tablehead{
 \colhead{Galaxy} & \colhead{Distance} & \colhead{$\Gamma_F$ or $\Gamma_S$} & \colhead{$\Sigma_{SFR}$} & \colhead{Age Range} &
 \colhead{M$_{min}$} & \colhead{CMF}  &
    \colhead{Spherically} & \colhead{Reference} \\
    \colhead{} & \colhead{(Mpc)} & \colhead{\%} & \colhead{($M_{\odot}$~yr$^{-1}$~kpc$^{-2}$)} & \colhead{(Myr)}  & \colhead{($M_{\odot}$)}  & \colhead{Shape}  & \colhead{Symmetric?} & \colhead{} } 
\startdata
 LMC & 0.05 & $5.8\pm0.5$ & $1.52\times10^{-3}$ & $10-100$  & 100 & Power law  & No & 1 \\
 SMC & 0.06 & $4.2^{+0.2}_{-0.3}$ & $6.83\times10^{-4}$ & $10-100$   & 100  & Power law  & No & 1 \\
M31 & 0.7 & $5.9\pm0.3$ & $2.34\times10^{-3}$ & $10-100$ & 100 & Schechter & No & 2 \\ 
IC10 & 0.7 & 4.2 & 0.03 & $1-10$  & 100 & Power law  & Yes & 3\\ 
NGC~1569 & 2.2 & $13.9\pm0.8$ & $2.8\times10^{-2}$ & $1-10$  & 100 & Power law &  No & 1 \\ 
NGC 7793 & 3.3 & $2.5\pm0.3$ & $6.43\times10^{-3}$ & $10-100$  & 10  & Schechter  & No & 4\\
NGC~4449 & 3.8 & 9 & 0.04 & $1-10$  & 1000   & Power law  & No & 5\\
NGC 4395 & 4.2 & $1.0\pm0.6$ & $4.66\times10^{-3}$ & $10-100$ & 300  & Schechter  & No & 4 \\
NGC 1313 & 4.4 & $3.2\pm0.2$ & $1.13\times10^{-2}$ & $10-100$  & 10 & Schechter  & No & 4 \\
 M83 & 4.5 & $18.2\pm3.0$ & $1.3\times10^{-2}$ & $1-10$  & 100  & Power law  & Yes & 6 \\
NGC 45 & 4.8 & $5.2\pm0.3$ & $1.01\times10^{-3}$ & $10-100$   & 10  & Schechter &  No & 4 \\
NGC~6946 & 5.5 & $12.5^{+1.8}_{-2.5}$ & $4.6\times10^{-3}$ & $1-10$ & 100 & Power law & No  & 1 \\
NGC~2997 & 9.6 & $10.0\pm2.6$ & $9.4\times10^{-3}$ & $1-100$  & 100  & Power law  & Yes & 7 \\
NGC 3256 & 36.0 & $22.9^{+7.3}_{-9.8}$ & 0.62 & $1-10$  & 100  & Power law  & No & 1 \\
ESO 338-IG04 & 37.5 & $50.0\pm10.0$ & 1.55 & $1-10$  & 100 & Power law  & No & 8 \\
SBS 0335-052E & 54.0 & $49.0\pm15.0$ & 0.95 & $1-10$  & 100  & Power law  & No & 8\\
Mrk 930 & 71.4 & $25.0\pm10.0$ & 0.59 & $1-10$ &  100 &  Power law  & No & 8 \\
ESO 185-IG13 & 76.3 & $26.0\pm5.0$ & 0.52 & $1-10$  & 100 & Power law  & No & 8 \\
Haro~11 & 82.3 & $50.0^{+13.0}_{-15.0}$ & 2.16 & $1-10$  & 100 &  Power law  & No & 8 \\ \hline
  \enddata
 \tablecomments{References: (1) Goddard et al. (2010); (2) Johnson et al. (2016); (3) Lim \& Lee (2015); (4) Silva-Villa \& Larsen 2011; (5) Annibali et al. (2011); (6) Adamo et al. (2015); (7) Ryon et al. (2014); (8) Adamo et al. (2011)  }
\end{deluxetable}
\end{rotatetable}


\begin{references}{}


\reference{ref1} Adamo, A., Ostlin, G., \& Zackrisson, E. 2011, MNRAS, 417, 1904

\reference{ref1} Adamo, A. et al. 2015 MNRAS, 452, 246

\reference{ref1} Annibali, F., Tosi, M., Aloisi, A., \& van der Marel, R. P. 2011, AJ, 142, 129

\reference{ref1} Bastian, N. 2008 MNRAS, 390, 759
  
\reference{ref1} Bastian, N., Adamo, A., Gieles, M., Silva-Villa, E., Lamers, H. J. G. L. M., Larsen, S. S., Smith, L. J., Konstantopoulos, I. S., Zackrisson, E. 2012, MNRAS, 419, 2606

 \reference{ref1} Bastian, N., Covey, K. R., \& Meyer, M. R. 2010, ARA \& A, 48, 339

 \reference{ref1} Baumgardt, H., \& Kroupa, P. 2007, MNRAS, 380, 1589

\reference{ref1} Binney, J., \& Tremaine, S. 2008, Galactic Dynamics, 2nd ed. (Printon, NJ: Princeton University Press)

\reference{ref1} Chabrier, G. 2003, PASP, 115, 763

 \reference{ref1} Chandar, R., Fall, S. M., \& Whitmore, B. C. 2015, ApJ, 810, 1 (CFW15)

\reference{ref1} Chandar, R., Whitmore, B. C., Calzetti, D., \& O'Connell, R. 2014, ApJ, 787, 17

\reference{ref1} Chandar, R., Fall, S. M., \& Whitmore, B. C. 2010a, ApJ, 711, 1263

\reference{ref1} Chandar, R., Whitmore, B. C., \& Fall, S. M. 2010b, ApJ, 713, 1343

\reference{ref1} Chandar, R., Whitmore, B. C., \& Kim, H. et al. 2010c, ApJ, 719, 966

\reference{ref1} Chandar, R., Whitmore, B. C., Dinino, D., et al. 2016, ApJ, 824, 71

\reference{ref1} Dalcanton, J. et al. 2012, ApJS, 200, 18


\reference{ref1} Fall, S. M., Chandar, R., \& Whitmore, B. C. 2005, ApJ, 631, L133

\reference{ref1} Fall, S. M., Chandar, R., \& Whitmore, B. C., 2009, ApJ, 704, 453

\reference{ref1} Fall, S. M., \& Chandar, R. 2012, ApJ, 752, 96

\reference{ref1} Fouesneau, M., Johnson, L. C., Weisz, D. et al. 2014, ApJ, 786, 117

\reference{ref1} Goddard, Q. E., Bastian, N. \& Kennicutt, R. C. 2010, MNRAS, 405, 857

\reference{ref1} Harris, J. \& Zaritsky, D. 2004, AJ, 127, 1531

\reference{ref1} Harris, J. \& Zaritsky, D. 2009, AJ, 138, 1243

\reference{ref1} Hunter, D., Elmegreen, B. G., Dupuy, T. J., \& Mortonson, M. 2003, AJ, 126, 1836

\reference{ref1} Indebetouw, R. et al. 2008 AJ, 136, 1442

\reference{ref1} Johnson, C. L., Seth, A. C., Dalcanton, J. J. et al. 2016, ApJ, 827, 33


\reference{ref1} Kennicutt, R. C. \& Evans, N. J. 2012, ARA\& A, 50, 531

\reference{ref1} Kroupa, P. 2001, MNRAS, 322, 231

\reference{ref1} Kruijssen, J. M. D. 2012, MNRAS, 426, 3008

\reference{ref1} Kruijssen, J. M. D. \& Bastian, N. 2016 MNRAS, 457, 24


\reference{ref1} Lada, C. J. \& Lada, E. A. 2003, ARA\&A, 41, 57

\reference{ref1} Lim, S. \& Lee, M. G. 2015 ApJ, 804, 123


\reference{ref1} Leitherer, C. Schaerer, D., \& Goldader, J. D. et al. 1999, ApJS, 123, 3

\reference{ref1} Leroy, A. et al. 2013, AJ 146, 19


\reference{ref1} McKee, C. F. \& Ostriker, E. C. 2007, ARA\&A, 45, 565


\reference{ref1} Messa, M. et al. 2017, MNRAS, in press (arXiv:1709.06101)

\reference{ref1} Mulia, A. J., Chandar, R., \& Whitmore, B. C. 2016, ApJ, 826, 32

\reference{ref1} Portegies Zwart, S. F., McMillan, S., \& Gieles, M. 2010 ARA\&A, 48, 431

\reference{ref1} Rangelov, B., Prestwich, A. H., \& Chandar, R. 2011, ApJ, 741, 86

\reference{ref1} Salpeter, E. 1955, ApJ, 121, 161

\reference{ref1} Silva-Villa, E., \& Larsen, S. S. 2011, A\& A, 529, 25

\reference{ref1} Silva-Villa, E., Adamo, A., Bastian, N., Fouesneau, M., \& Zackrisson, E. 2014, MNRAS, 440, 116

\reference{ref1} Spitzer, L Jr. 1958, ApJ, 127, 544

\reference{ref1} Whitmore, B. C., Chandar, R., \& Fall, S. M. 2007, AJ, 133, 1067

\reference{ref1} Whitmore, B.C. \& Zhang, Q.  2002, AJ, 124, 1418

\reference{ref1} Whitmore, B.C., Chandar, R., Schweizer, F. et al. 2010, AJ, 140, 75

\reference{ref1} Williams, B. et al. 2014 ApJS, 215, 9

\reference{ref1} Zhang, Q., Fall. S. M., \& Whitmore, B. C. 2001, ApJ, 561, 727


\end{references}
\end{document}